\documentclass[iop,twocolappendix]{emulateapj}

\usepackage{epsfig}

\usepackage{amssymb}

\usepackage{multirow}

\usepackage{enumitem}

\usepackage{ulem}

\usepackage{rotating}
\usepackage{natbib}
\usepackage{latexsym}
\bibpunct{(}{)}{;}{a}{}{,}
\usepackage{url}

\begin{document}

\newcommand{\Msun}{M$_\mathrm{\odot}$}

\def\rem#1{{\bf #1}}
\def\hide#1{}

\def \aj {AJ}
\def \mnras {MNRAS}
\def \apj {ApJ}
\def \apjs {ApJS}
\def \apjl {ApJL}
\def \aap {A\&A}
\def \aapr {A\&ARv}
\def \nat {Nature}
\def \araa {ARAA}
\def \pasp {PASP}
\def \aaps {AAPS}
\def \prd {PhRvD}
\def \apss {ApSS}

\newcommand{\specialcell}[2][c]{%
  \begin{tabular}[#1]{@{}c@{}}#2\end{tabular}}

\title{Peering into the dark side: magnesium lines establish a massive
  neutron star in PSR~J2215+5135}

\author{M. Linares\altaffilmark{1,2}, T. Shahbaz\altaffilmark{2,3},
  J. Casares\altaffilmark{2,3}}
%


\submitted{The Astrophysical Journal, in press.}

\altaffiltext{1}{Departament de F{\'i}sica, EEBE, Universitat Polit{\`e}cnica de Catalunya, c/ Eduard Maristany 10, 08019 Barcelona, Spain}

\altaffiltext{2}{Instituto de Astrof{\'i}sica de Canarias, c/ V{\'i}a
  L{\'a}ctea s/n, E-38205 La Laguna, Tenerife, Spain}
\altaffiltext{3}{Universidad de La Laguna, Departamento de
  Astrof{\'i}sica, E-38206 La Laguna, Tenerife, Spain}
%


\vspace{0.5cm}

\keywords{pulsars: general --- pulsars: individual(PSR~J2215+5135) ---
  stars: neutron --- binaries: general --- stars: variables: general
  --- X-rays: binaries}

\begin{abstract}

New millisecond pulsars (MSPs) in compact binaries provide a good
opportunity to search for the most massive neutron stars.
Their main-sequence companion stars are often strongly irradiated by
the pulsar, displacing the effective center of light from their
barycenter and making mass measurements uncertain.
We present a series of optical spectroscopic and photometric
observations of PSR~J2215+5135, a ``redback'' binary MSP in a 4.14~hr
orbit, and measure a drastic temperature contrast between the dark/cold
(T$_\mathrm{N}$=5660$^{+260}_{-380}$~K) and bright/hot
(T$_\mathrm{D}$=8080$^{+470}_{-280}$~K) sides of the companion star.
We find that the radial velocities depend systematically on the
atmospheric absorption lines used to measure them.
Namely, the semi-amplitude of the radial velocity curve of J2215
measured with magnesium triplet lines is systematically higher than
that measured with hydrogen Balmer lines, by 10\%.
We interpret this as a consequence of strong irradiation, whereby
metallic lines dominate the dark side of the companion (which moves
faster) and Balmer lines trace its bright (slower) side.
Further, using a physical model of an irradiated star to fit
simultaneously the two-species radial velocity curves and the
three-band light curves, we find a center-of-mass velocity of
K$_2$=412.3$\pm$5.0~km~s$^{-1}$ and an orbital inclination
i=63.9$^\circ$$^{+2.4}_{-2.7}$.
  Our model is able to reproduce the observed fluxes and velocities
without invoking irradiation by an extended source.
We measure masses of M$_1$=2.27$^{+0.17}_{-0.15}$~M$_\odot$ and
M$_2$=0.33$^{+0.03}_{-0.02}$~M$_\odot$ for the neutron star and the
companion star, respectively.
If confirmed, such a massive pulsar would rule out some of the
proposed equations of state for the neutron star interior.

\vspace{0.5cm}

\end{abstract}

\maketitle

\section{Introduction}
\label{sec:intro}


New millisecond pulsars (MSPs) in compact binaries (orbital period
P$_\mathrm{orb}$$\lesssim$1~d)
are being discovered with the advent of the {\it Fermi} large area
telescope \citep[LAT;][]{Atwood09}.
Their companion or secondary stars are light ($\sim$0.1~M$_\odot$
in the so-called {\it redbacks}) or ultralight
($\sim$0.01~M$_\odot$ in the {\it black widows}),
and in some cases they are strongly irradiated by the pulsar wind and
high-energy radiation powered by the neutron star's rotational energy
loss ($\dot{E}$).
Furthermore, three of the nearly twenty known redback MSPs have shown
transitions between the radio-pulsar and accretion-disk states, which
has provided a long-sought link between MSPs and low-mass X-ray
binaries (LMXBs; \citealt{Archibald09,deMartino13,Papitto13b}; see
\citealt{Linares14c} for a review of redback states).


Most of these new compact binary MSPs are relatively nearby
($\lesssim$4~kpc) and far from the Galactic plane
($\gtrsim$5$^\circ$), where interstellar extinction is low.
This allows for sensitive optical spectroscopic observations
and dynamical studies of the companion star in its orbit around the
pulsar, and offers a new opportunity to measure the mass of spun-up
``recycled" neutron stars \citep[e.g.,][]{Romani11,Kaplan13,Crawford13}.
However, as we discuss in the present work, the effects of
irradiation on the measured radial velocities must be carefully
taken into account in order to avoid systematic uncertainties.


PSR~J2215+5135 (J2215 hereafter) was discovered as a 2.61-ms MSP in
radio searches of the LAT source 1FGL~J2216.1+5139 (i.e.,
2FGL~J2215.7+5135 or 3FGL~J2215.6+5134), and to date has the shortest
P$_\mathrm{orb}$ among Galactic field redbacks
\citep[P$_\mathrm{orb}$$\simeq$4.14~hr;][]{Hessels11}.
Even though this system has been observed so far only in the
(disk-free, rotation powered) {\it pulsar state}, \citet{Linares14c}
found a relatively high X-ray luminosity
L$_\mathrm{X}$$\sim$10$^{32}$~erg~s$^{-1}$,
suggesting J2215 as a candidate for future transitions to an accreting
state.
Optical photometry revealed a V=20.2--18.7 mag counterpart with
orbital variability typical of strongly irradiated systems
\citep{Breton13,Schroeder14}.
Modelling the optical lightcurves (LCs) can determine the
inclination of the orbit which, together with the precise 
ephemerides obtained from pulsar timing, may allow a full orbital
solution and a neutron star mass measurement (e.g.,
\citealt{Kerkwijk11}; see also \citealt{Shahbaz98,Casares06}).
Nevertheless, the orbital parameters for J2215 presented by
\citet[based on LC model fits]{Schroeder14} and by \citet[including
  also radial velocities]{Romani15} differ by a large amount, yielding
inconsistent neutron star masses M$_\mathrm{NS}$ in the range
1.6-2.5~M$_\odot$.
%


We present here the results of a new set of observations of J2215
taken in 2014 with three different telescopes
(Section~\ref{sec:data}), including the 10.4-m {\it Gran Telescopio
  Canarias} (GTC).
These reveal an extreme temperature contrast between the cold/dark
(``night'') and hot/bright (``day'') faces of the secondary star
(Section~\ref{sec:temp}).
In order to place tighter independent constraints on M$_\mathrm{NS}$
and to investigate systematic effects on dynamical studies of this new
class of pulsars, we carefully measure the spectral type and radial
velocity of the companion along the orbit (Sec.~\ref{sec:molly}).

\begin{table*}[ht]
\scriptsize
\caption{Summary of optical observations of PSR~J2215+5135.}
\begin{minipage}{\textwidth}
\centering
\begin{tabular}{c c c c c c c c c}
\hline\hline
%
Telescope & Instrument & Band\footnote{Effective wavelength of the photometric filters or approximate wavelength range covered by the spectra, in Angstroms.} & Date & Time & Exposures & Orbital & Airmass & Seeing \\
(diameter) & (configuration) & ($\lambda$, \AA) & (evening) & (UT) & (nr.$\times$duration) & phase & & ('')\\
\hline
\multicolumn{9}{c}{\textbf{Photometry}}\\
\hline
IAC-80 & CAMELOT & g',r' & 2014-08-02 & 22:53-04:01 & 4$\times$420s & 0.4-1.6 & 1.47-1.09 & 1.1-2.5\\
(80~cm) & (bin 2$\times$2) & (4639,6122)  &  &  & +26$\times$600s &  &  & \\
\hline
IAC-80 & CAMELOT & g',r' & 2014-08-03 & 23:32-03:52 & 30$\times$420s & 0.3-1.3 & 1.32-1.09 & 0.4-0.7 \\
(80~cm) & (bin 2$\times$2) & (4639,6122)  &  &  &  &  &  & \\
\hline
WHT & ACAM & g',r',i' & 2014-08-11 & 22:15-00:03 & 95$\times$60s & 0.4-0.8 & 1.48-1.18 & 0.7-1.6 \\
(4.2~m) & (win. 501$\times$501) & (4639,6122,7439)  &  &  &  &  &  & \\
\hline
WHT & ACAM & g',r',i' & 2014-09-01 & 23:17-02:06 & 149$\times$60s & 0.4-1.0 & 1.13-1.09 & 0.8-1.4 \\
(4.2~m) & (win. 501$\times$501) & (4639,6122,7439)  &  &  &  &  &  & \\
\hline\hline
\multicolumn{9}{c}{\textbf{Spectroscopy}}\\
\hline
WHT & ISIS & B:3700-5300 & 2014-08-11 & 00:23-05:19 & 17$\times$900s & 0.0-1.1 & 1.17-1.33 & 0.7-1.6 \\
(4.2~m) & (R600B+R) & R:5500-7200  &  &  &  &  &  &  \\
\hline
GTC & OSIRIS & B:4000-5700 & 2014-11-14 & 19:47-21:06 & 5$\times$935s & 0.6-0.8 & 1.09-1.11 & 0.6-1.1 \\
(10.4~m) & (R2000B) &  &  &  &  &  &  & \\
\hline
GTC & OSIRIS & B:4000-5700 & 2014-11-15 & 19:43-23:58 & 16$\times$935s & 0.3-1.3 & 1.09-1.55 & 0.7-1.3 \\
(10.4~m) & (R2000B) &  &  &  &  &  \\
\hline\hline
\end{tabular}
\end{minipage}
\label{table:data}
\end{table*}

We find that the apparent radial velocities of J2215 depend on both
the spectral range and the reference/template spectrum used to measure
them (Section~\ref{sec:rvcs}).
In Section~\ref{sec:model}, we model jointly the observed LCs and
radial velocities, including for the first time dynamical information
of the cold/dark side of the companion.
We find a new orbital solution (Sec.~\ref{sec:modres}) with an
extremely massive neutron star (Sec.~\ref{sec:mass}).
We discuss these results in Section~\ref{sec:discussion}, as well as
the implications for dynamical studies in compact binaries with
strong irradiation.
Section~\ref{sec:conclusions} contains a summary of our main results
and conclusions.

\begin{figure}[h]
  \begin{center}
  \resizebox{1.0\columnwidth}{!}{\rotatebox{0}{\includegraphics[]{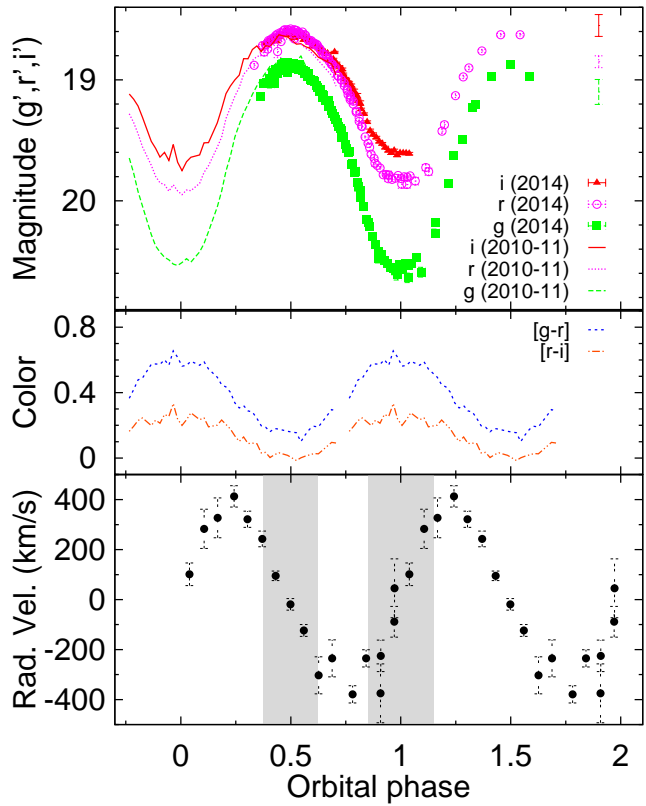}}}
%
  \caption{ 
{\it Top:} Optical light curves of J2215 in three bands, as indicated. Data points show our 2014 IAC-80 \& WHT observing campaign and lines (solid, dotted and dashed) show the 2010-2011 results from \citet{Schroeder14}. Error bars on the top right corner show the uncertainty on the magnitude calibration (Sec.~\ref{sec:photo}; errors on differential magnitude are plotted but smaller than the symbols).
{\it Middle:} Color variation along the orbit showing redder/colder emission around light minimum (orbital phase 0).
{\it Bottom:} Radial velocity curve from our WHT-ISIS spectra, calculated by cross-correlating the full spectra (red and blue arms) with an F5 template. The averages used for optimal subtraction are shown with gray-shaded rectangles (Sec.~\ref{sec:molly}).
} %
    \label{fig:rvlc}
 \end{center}
\end{figure}

\section{Observations, reduction and analysis}
\label{sec:data}

\subsection{Photometry}
\label{sec:photo}

%
We obtained phase-resolved photometric observations of PSR~J2215+5135
with the IAC-80 and William Herschel (WHT) telescopes at the Teide and
Roque de los Muchachos observatories, respectively, on four different
nights (see Table~\ref{table:data}).
The IAC-80 images were taken on 2014 August 2--3 with the
Teide observatory light improved camera (CAMELOT; 0.30 arcsec
pixel$^{-1}$) using the SDSS filters g' and r', an exposure time of
either 420s or 600s and a binning of 2$\times$2 pixels in order to
optimize the signal-to-noise ratio (S/N).
The WHT images were taken on 2014 August 11 (contemporaneous with our
WHT-ISIS spectra) and September 1, with the auxiliary-port camera
(ACAM; 0.25 arcsec pixel$^{-1}$ in imaging mode) using the SDSS
filters g', r' and i', an exposure time of 60s and a window of
501$\times$501 pixels around the source in order to reduce readout
time (readout and filter change resulted in deadtime of 6-9s per
exposure).

All images were debiased and flat-fielded using standard {\sc iraf}
routines.
We then performed aperture photometry using the {\sc ultracam}
pipeline, a variable source extraction radius (set to 1.5--1.7 times
the seeing) and a nearby stable non-saturated reference star.
The resulting differential magnitudes are relative to a nearby star in
the field and thus insensitive to thin clouds or moderate atmospheric
variability (yet the observing conditions were generally good).
The absolute flux (apparent magnitude) calibration was done using a
nearby AAVSO-APASS star, with uncertainties of 0.05 mag (r') and 0.1
mag (g', i'), and checked against other nearby stars from the USNO-B1
catalog.
We also compared these reference star magnitudes with those given by
the PANSTARRS catalog, and found only a significant difference in the
r' band, with a shift of +0.16~mag with respect to the APASS values
that we use.
We verified that this has no impact on any of the results reported in
this work; in particular, the parameters reported in
Section~\ref{sec:model} are all consistent within the errors when
using the PANSTARRS instead of the APASS calibration.
Figure~\ref{fig:rvlc} (top) shows the J2215 LCs folded at the
orbital period.

\subsection{Spectroscopy}
\label{sec:spec}

We observed J2215 with the WHT and GTC telescopes on 2014 August 11
and November 14--15, respectively, in order to obtain medium
resolution spectra of the companion star and measure its spectral type
and velocity along the orbit.
For the WHT-ISIS spectra we chose the R600B (blue arm) and R600R (red
arm) gratings centered at 4500~\AA\ and 6400~\AA, respectively. The slit
width was set to 1'', resulting into a resolution of
105--130~km~s$^{-1}$ (R$\sim$2600) and 65--80~km~s$^{-1}$
(R$\sim$4000) for the blue and red arms, respectively.
At the GTC we used OSIRIS in its long-slit spectroscopy mode, with the
R2000B VPH gratings and a slit width of 1'', resulting into a
resolution of 145--160 km~s$^{-1}$ (R$\sim$2000).
With this campaign we obtained 17 WHT-ISIS and 21 GTC-OSIRIS spectra
covering the full 4.14-hr orbit with some redundancy and with exposure
times of 900 s and 935 s, respectively (i.e., exposing each spectrum
for about 6\% of the J2215 orbit; see Table~\ref{table:data}).

After applying bias and flat corrections to the trimmed images within
{\sc iraf}, we extracted the spectra and subtracted sky background
using the optimal extraction method within {\sc starlink/pamela} to
account for the significant tilt \citep{Marsh89}.
The WHT-ISIS spectra were calibrated in wavelength using interspersed
arc spectra (CuNe, CuAr) extracted from the same source extraction
regions, taken once every two source spectra.
A set of well identified arc lines were satisfactorily fitted with a
4th order polynomial to produce the wavelength scale (47 and 31 lines
in the blue and red arms, resulting into an rms amplitude of residuals
of 0.05~\AA\ and 0.02~\AA, respectively).
We adjusted the same polynomial function to all arc spectra and
interpolated in time between adjacent arcs to calibrate the science
spectra, thereby accounting for the significant ($\sim$1~\AA) drift
due to instrument flexure.
The GTC-OSIRIS wavelength calibration was done using one set of arcs
taken on the second night, fitting the pixel-wavelength relation with
a 4th order polynomial (19 lines giving residuals with an rms
amplitude of 0.06~\AA).
The resulting wavelength calibration was checked and refined using the
OI sky line at 5577~\AA, which allowed us to correct for residual
($\lesssim$10~km~s$^{-1}$) shifts in the wavelength solution.

\begin{figure*}[ht]
  \begin{center}
  \resizebox{1.75\columnwidth}{!}{\rotatebox{-90}{\includegraphics[]{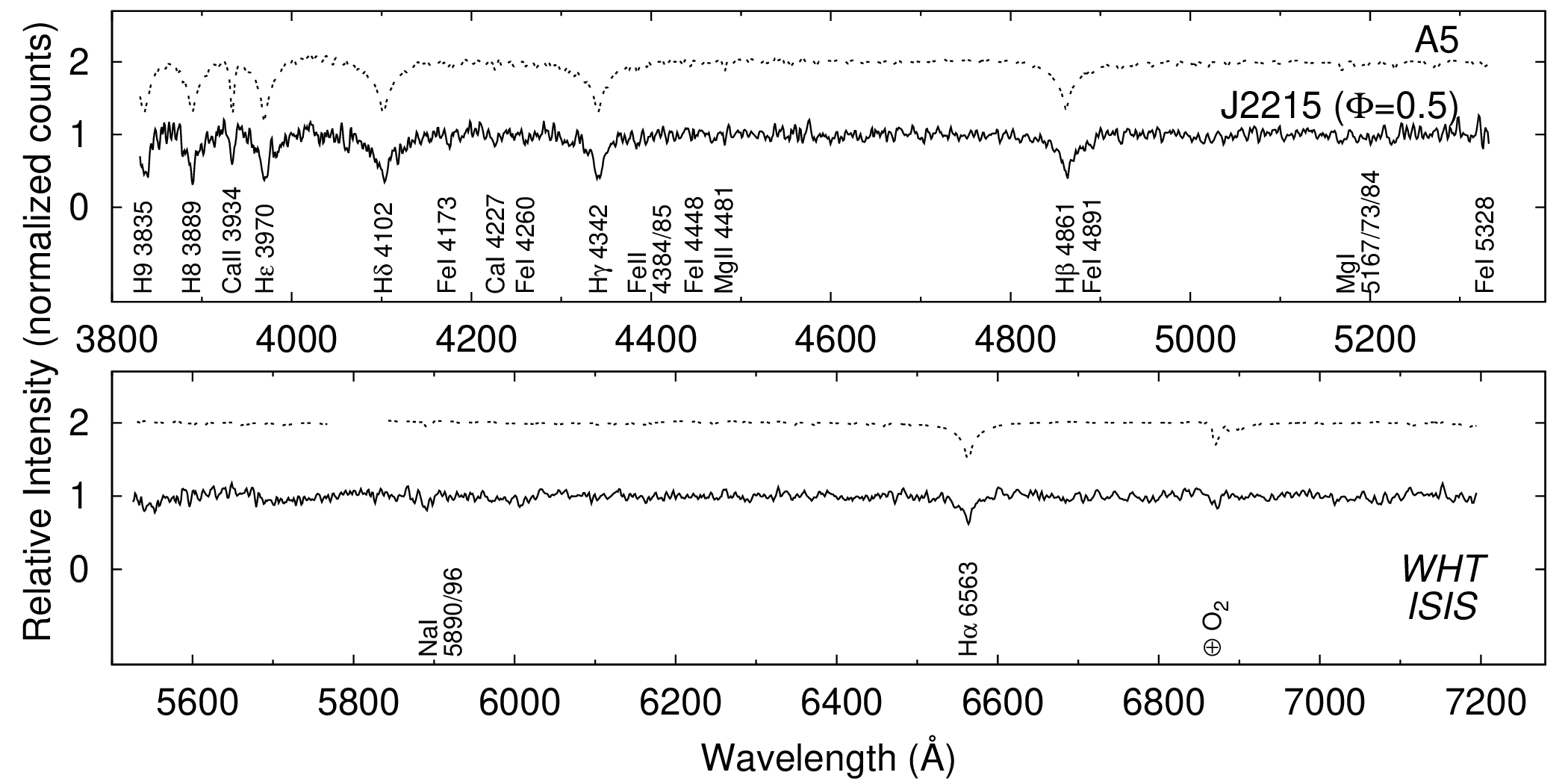}}}
  \caption{
Normalized WHT-ISIS spectra of {\it (from bottom up, arbitrary shifts
  for display purpose)} J2215 around orbital phase 0.5 (solid line)
and an A5 spectral type standard star (dotted line).
The main absorption lines are identified along the bottom axes.
} %
  \label{fig:isis}
  \resizebox{1.75\columnwidth}{!}{\rotatebox{-90}{\includegraphics[]{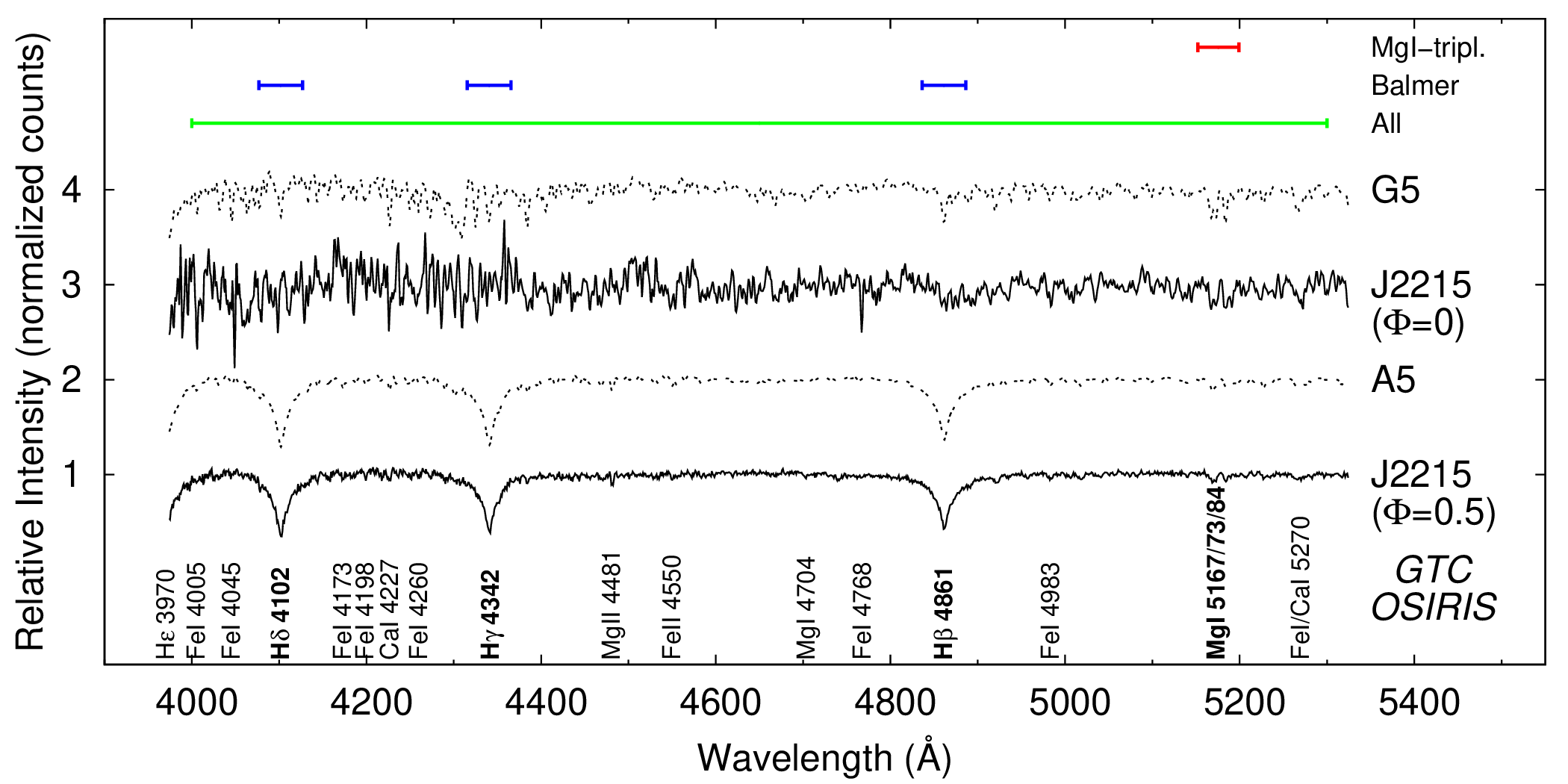}}}
  \caption{
Normalized GTC-OSIRIS spectra of {\it (from bottom up, arbitrary
  shifts for display purpose)}: J2215 around orbital phase 0.5 (i.e.,
companion at superior conjunction; solid line); an A5 spectral type
standard star (dotted line); J2215 around orbital phase 0 (companion
at inferior conjunction; solid line) and a G5 spectral type standard
star (dotted line).
The main absorption lines are identified along the bottom axis.
Green/blue/red lines along the top axis show the ranges used for
radial velocity measurements (Sec.~\ref{sec:molly}).
} %
    \label{fig:osiris}
 \end{center}
\end{figure*}

\subsection{Spectral analysis}
\label{sec:molly}

%

In order to measure the radial velocity and temperature of the
companion star in J2215 throughout the orbit, we applied within {\sc
 molly} the {\it cross correlation} and {\it optimal subtraction}
techniques, respectively.
As both techniques require comparison spectra (or templates
hereafter), we built a library of main-sequence stellar templates with
spectral types between O4 and M0 (see Appendix~\ref{app:uvespop}).
The continuum level of all 38 source and 33 template spectra was
normalized with a spline fit, and subtracted.

After binning to a same heliocentric velocity scale, excluding
telluric lines and broadening the template spectra to the source
spectral resolution (Sec.~\ref{sec:spec}), each source and template
spectra were cross correlated to find their relative velocity,
allowing shifts between -700 and +700 km~s$^{-1}$.
The resulting radial velocity curves (RVCs) from the GTC and WHT
spectra were fitted with a sine function
V=G+Ksin[2$\pi$(t-T$_0$)/P$_\mathrm{orb}$], where V and G are
the radial and systemic velocities, K is the semi-amplitude of the
RVC, t is the middle time of each spectrum, T$_0$ is the time of
inferior conjunction of the secondary which defines\footnote{Note
  different definition than usual pulsar timing phase zero, which
  marks the time of ascending node of the pulsar.} orbital phase
$\Phi_\mathrm{orb}$=0 and P$_\mathrm{orb}$ is the orbital period.
Having checked that the best-fit period is consistent with (but less
precise than) the orbital period from pulsar timing \citep{Abdo13},
this parameter was subsequently fixed at the pulsar timing value
P$_\mathrm{orb}$=0.1725021049[8]~d.

Using the orbital parameters above, we corrected for the systemic
velocity and orbital motion and shifted all source spectra to the
reference frame of each template.
In order to increase the S/N, we averaged 4--6 source spectra around
$\Phi_\mathrm{orb}$=0$\pm$0.15 and $\Phi_\mathrm{orb}$=0.5$\pm$0.125.
We then performed an optimal subtraction using the full GTC spectral
range, i.e., subtracted the templates scaled by a factor
f$_\mathrm{veil}$ from the source averaged spectra, adjusting
f$_\mathrm{veil}$ to minimize the residual scatter.
This is a quantitative way of matching the observed absorption lines
from J2215 to a set of templates with known spectral types and
temperatures \citep[]{Marsh94}.

Because J2215 becomes very faint around $\Phi_\mathrm{orb}$=0
(r$\simeq$20~mag; Fig.~\ref{fig:rvlc}, where the cold face of the
companion star dominates), the corresponding ISIS spectra have low
S/N.
The tightest constraints on the spectral type and radial velocities
come from the higher S/N GTC spectra, so we focus the rest of our
analysis on those.
Only one out of the 21 GTC spectra could not be included in the
analysis due to the very low ($<$100 at peak) number of counts
collected.

Thanks to the large GTC collecting area we were able to measure radial
velocities using i) the {\it full spectral range} (4000--5300~\AA),
ii) the hydrogen {\it Balmer lines} (three 50 \AA-wide windows
centered on H$\beta$, H$\gamma$ and H$\delta$) and iii) the {\it Mg-I
  triplet} lines present in the J2215 spectra (5152--5199~\AA).
The corresponding spectral ranges and two representative GTC-OSIRIS
spectra are shown in Figure~\ref{fig:osiris}.
As the absorption spectra of early/hot and late/cold stars are
dominated by different sets of lines (see, e.g.,
Appendix~\ref{app:uvespop}), the Balmer and Mg-I radial velocities
allow us to track different parts of the irradiated companion
throughout its orbit around the pulsar.

The cross correlation of two broad lines may yield ambiguous results
if the profiles are not exactly the same. The matching wings can give
two maxima in the cross correlation function, at velocities which
differ from that measured using the line core (``double peaked'' cross
correlation functions).
On close inspection of J2215's RVCs, we found that this introduces
strong artificial deviations from a sinusoidal function around phase
0.5, only when using Balmer lines {\it and} templates of spectral type
earlier than F (i.e., when both source and template spectra are
dominated by broad lines).
For this reason, we include only spectral types later than F0 in our
results for the Balmer-line RVCs (Section~\ref{sec:rvcs}).

\section{Results}
\label{sec:results}

\subsection{Temperature of the hot and cold sides}
\label{sec:temp}


The optical flux from J2215 varies smoothly along the orbit (no flares
are detected in the 60s cadence data), with one clear maximum and
minimum per orbital cycle and a peak-to-peak amplitude of almost two
magnitudes (g', B bands).
We find that the orbital LCs (Fig.~\ref{fig:rvlc}) are stable
over timescales of years, comparing our 2014 observations with the
2010-2011 data presented by \citet[][converting their BVR magnitudes
into the SDSS g'r'i' system following \citealt{Jordi06}]{Schroeder14}.
The $\sim$0.4~mag change in the [g'-r'] color reveals hotter emission
at maximum light ($\Phi_\mathrm{orb}$=0.5). 
The RVC of the companion shows a large amplitude
(K$\sim$400~km~s$^{-1}$) and changes sign near the maximum and minimum
light ($\Phi_\mathrm{orb}$=0.5
and 0, respectively), as can be seen already from the WHT spectra
(Fig.~\ref{fig:rvlc}) and as first reported by \citet{Romani15}.
These properties are indicative of a companion star that is strongly
irradiated by the pulsar throughout the compact 4.14~hr orbit.

\begin{figure}[h]
  \begin{center}
  \resizebox{1.0\columnwidth}{!}{\rotatebox{-90}{\includegraphics[]{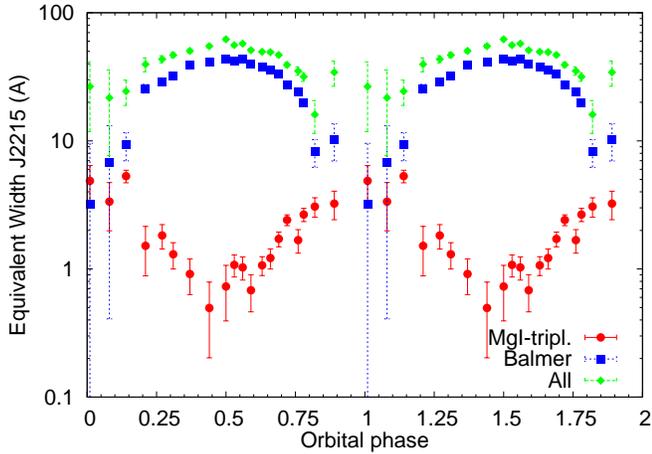}}}
  \caption{
Equivalent width (EW) of all (green diamonds), Balmer (blue squares)
and Mg-I (red circles) absorption lines in the GTC spectra of J2215,
as a function of orbital phase.
} %
    \label{fig:ewsrc}
 \end{center}
\end{figure}

\begin{figure}[ht]
  \begin{center}
  \resizebox{1.0\columnwidth}{!}{\rotatebox{-90}{\includegraphics[]{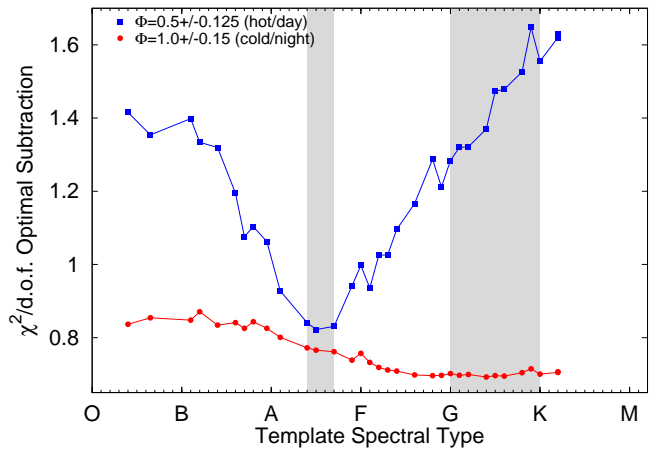}}}
%
  \caption{
Reduced chi squared of optimal subtraction plotted vs. template
spectral type.
Minima show the spectral type of the optical companion to J2215 around
orbital phase 0.5 (blue squares, left shaded region) and phase 1 (red
circles, right shaded region; see Sec.~\ref{sec:molly}).
} %
    \label{fig:phst}
 \end{center}
\end{figure}

\begin{figure}[h]
  \begin{center}
  \resizebox{1.0\columnwidth}{!}{\rotatebox{-90}{\includegraphics[]{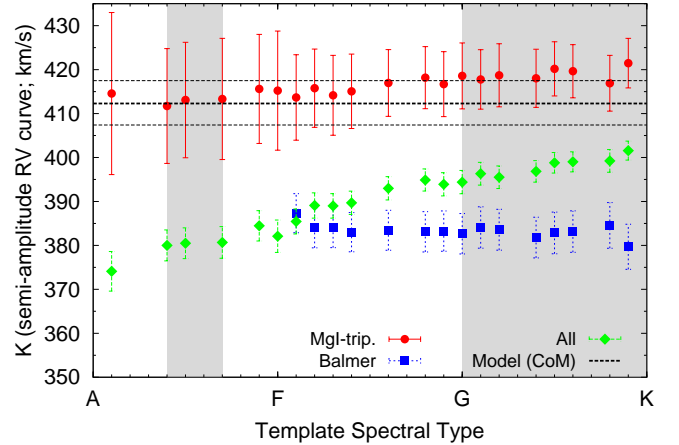}}}
%
  \caption{
%
%
Semi-amplitude of the RVC of J2215 as a function of
template spectral type, calculated using all absorption lines (green
diamonds), MgI lines (red circles) and Balmer lines (blue squares).
Shaded regions show the spectral type of the hot/cold sides, and
horizontal lines show the value of the center of mass
semi-amplitude from our model (Sec.~\ref{sec:model}).
} %
    \label{fig:k2}
 \end{center}
\end{figure}

\begin{figure*}[ht]
  \begin{center}
  \resizebox{1.0\columnwidth}{!}{\rotatebox{-90}{\includegraphics[]{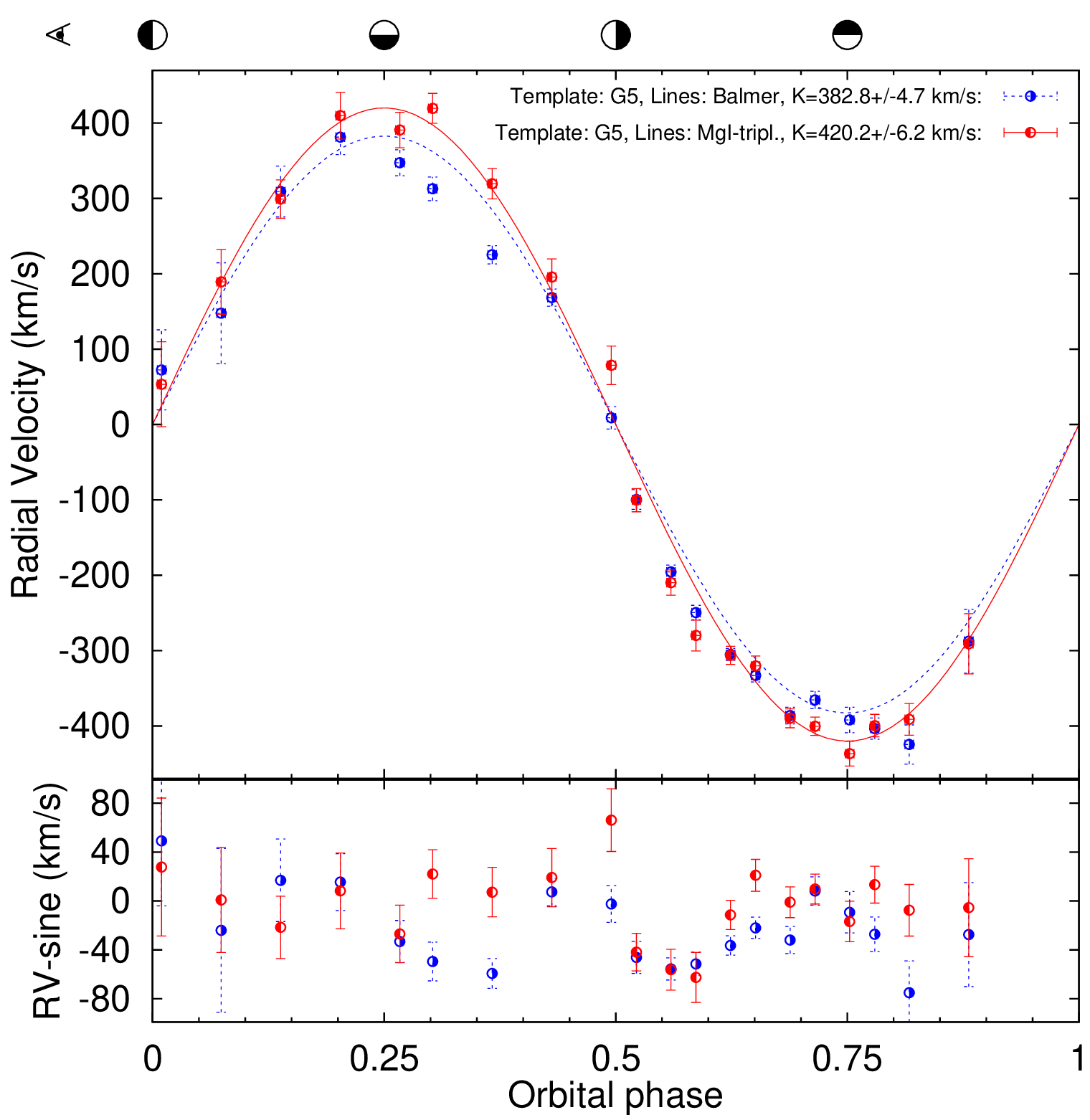}}}
  \resizebox{1.0\columnwidth}{!}{\rotatebox{-90}{\includegraphics[]{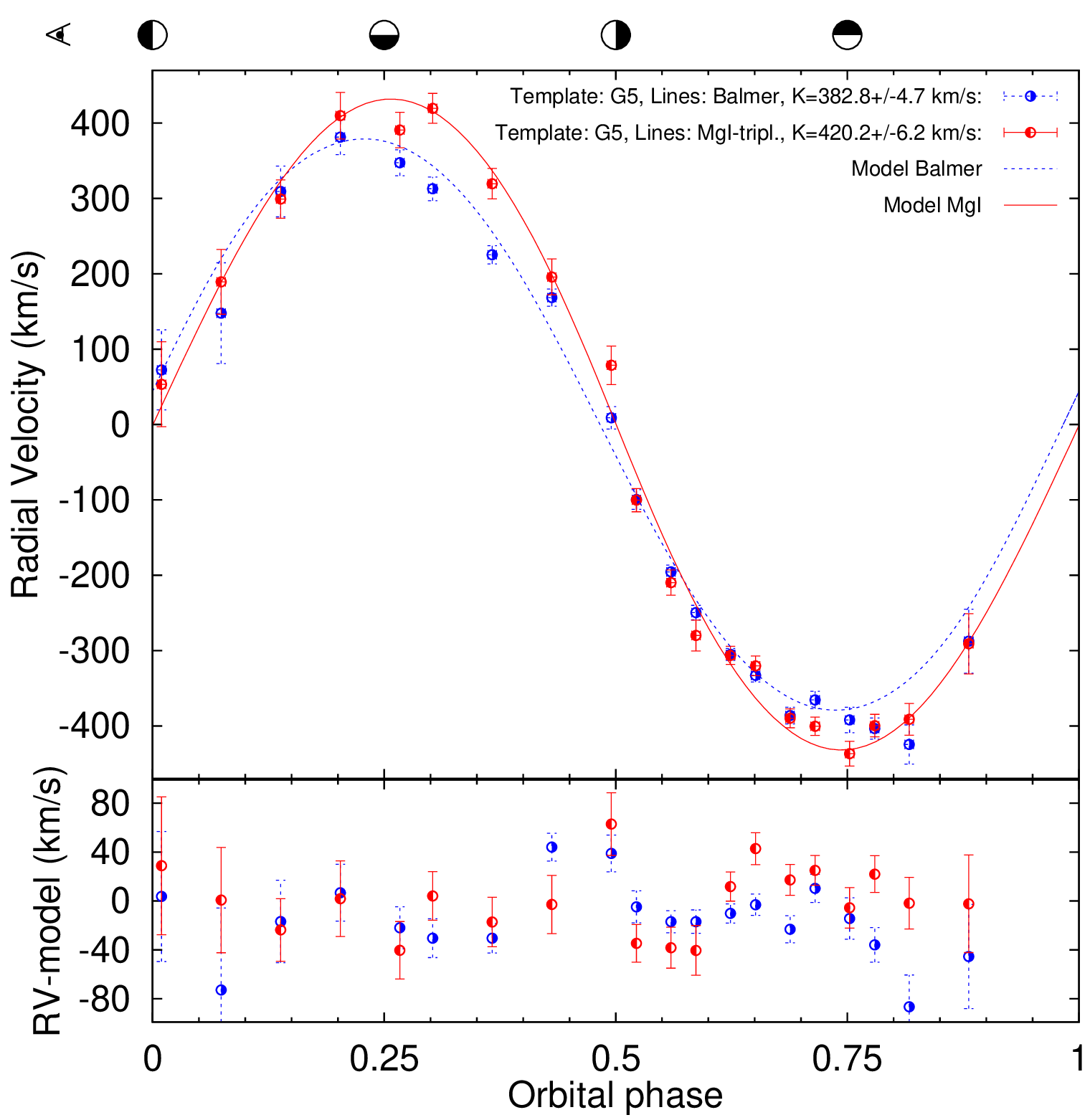}}}
  \caption{
Radial velocity of the companion star in J2215 in its orbit around the
pulsar, as measured by cross-correlation with a G5 template using i)
Balmer lines (blue points) and ii) MgI triplet lines (red points;
Sec.~\ref{sec:rvcs}).
{\it Left: } Blue dashed and red solid lines show the sinusoidal fits
to the Balmer and MgI RVCs, respectively (the best-fit systemic
velocity was subtracted in both cases).
Sine fit residuals are shown in the bottom panel.
{\it Right: } Best fit and residuals from our {\sc xrbcurve} model of a
nearly Roche lobe filling irradiated star (Sec.~\ref{sec:model}).
  } %
    \label{fig:rvcompare}
 \end{center}
\end{figure*}


The J2215 spectra around $\Phi_\mathrm{orb}$=0.5
(Figures~\ref{fig:isis} and \ref{fig:osiris}) show strong Balmer lines
(H1 through H9) consistent with an A5 star, as well as numerous yet
weaker metallic (Ca/Fe/Mg) lines.
Around $\Phi_\mathrm{orb}$=0, when the companion star presents its
cold face to the observer, Balmer lines are much weaker and narrower
while Mg-I triplet lines are stronger.
Hence the equivalent widths (EW) of Balmer and Mg-I lines are
anticorrelated along the orbit, as shown in Figure~\ref{fig:ewsrc}.
The optimal subtraction analysis (Sec.~\ref{sec:molly}) gives a clean
measurement of the temperature and spectral type of the companion
star, independent of the measured colors (which may be contaminated by
non-stellar light).
We find drastic changes between the irradiated and cold sides of the
companion star.
This is shown qualitatively in Figures~\ref{fig:osiris} and
\ref{fig:isis}, where the J2215 GTC and WHT spectra are compared to A5
and G5 templates degraded to the same resolution.

Figure~\ref{fig:phst} shows our quantitative results: the reduced chi
squared resulting from the optimal subtraction method
(Sec.~\ref{sec:molly}) for templates with a broad range of spectral
types (O--M).
We thereby measure a spectral type A5$\pm$2 for the brightest spectra
(i.e., A3-A7 at $\Phi_\mathrm{orb}$=0.5) and G5$\pm$5 at the faintest
end (G0-K0 at $\Phi_\mathrm{orb}$=0; \citealt{Romani15} report similar
yet slightly earlier spectral types of A2 and G0 around phase 0.5 and
0, respectively, but no errors are quoted).
These correspond to effective temperatures for the cold (``night'')
and hot (``day'') sides of T$_\mathrm{N}$=5660$^{+260}_{-380}$~K and
T$_\mathrm{D}$=8080$^{+470}_{-280}$~K, respectively \citep{Pecaut13}.
We thus find a drastic temperature contrast between opposite sides of
the companion star, where the hot/day side is about 2400~K or 40\%
hotter than it would be without irradiation.
The best match scale factors are f$_\mathrm{veil}$$\simeq$0.8
at both superior and inferior conjunction, suggesting a contribution
from non-stellar light (veiling) of about 20\% in this GTC-OSIRIS
4000--5300~\AA\ band (which corresponds approximately to filter g').

\subsection{Radial velocities: magnesium vs. Balmer lines}
\label{sec:rvcs}


%
We find that the radial velocities and K values depend systematically
on the set of lines or {\it spectral range} used to measure them.
Namely, as we show in Figure~\ref{fig:k2}, the semi-amplitude of the
Mg-I line RVC (K$_\mathrm{Mg}$, red circles) is always $\sim$10\%
higher than the semi-amplitude of the Balmer line RVC
(K$_\mathrm{Balmer}$, blue squares).
Using the same G5 template yields
K$_\mathrm{Balmer}$=382.8$\pm$4.7~km~s$^{-1}$ and
K$_\mathrm{Mg}$=420.2$\pm$6.2~km~s$^{-1}$.

In Figure~\ref{fig:rvcompare} we present two extreme cases: RVCs of
J2215 calculated using {\it Balmer} lines (blue symbols) and {\it
  MgI-triplet} lines (red symbols; see Sec.~\ref{sec:molly} for the
exact wavelength ranges). In both cases the radial velocities were
measured by cross correlating the J2215 spectra with the same G5
template.
The MgI triplet lines yield higher radial velocities than the Balmer
lines (in absolute value) around orbital phases 0.25 and 0.75, when
the companion star is viewed sideways (in ``quadrature'').
Because the velocities at quadrature are extreme, this has an
important impact on the measured K values and the corresponding
neutron star mass (Sec.~\ref{sec:Kcorr}).
Most RVCs are reasonably well fitted with a pure sinusoidal function
(reduced chi-squared $\lesssim$2). There are, however, deviations
noticeable in the sine fit residuals (Fig.~\ref{fig:rvcompare}),
especially in the Balmer line RVCs (spectral types earlier than F1
were not used in the Balmer line cross correlation analysis, see
Sec.~\ref{sec:molly}).
The fitted K values were verified in a model independent way by
measuring the peak to peak semi-amplitude of the RVC: the two maximum
and minimum radial velocities were averaged, subtracted and divided by
two. The results were always consistent with the K values presented in
Figure~\ref{fig:k2}.

\begin{figure}[h]
  \begin{center}
  \resizebox{1.\columnwidth}{!}{\rotatebox{0}{\includegraphics[]{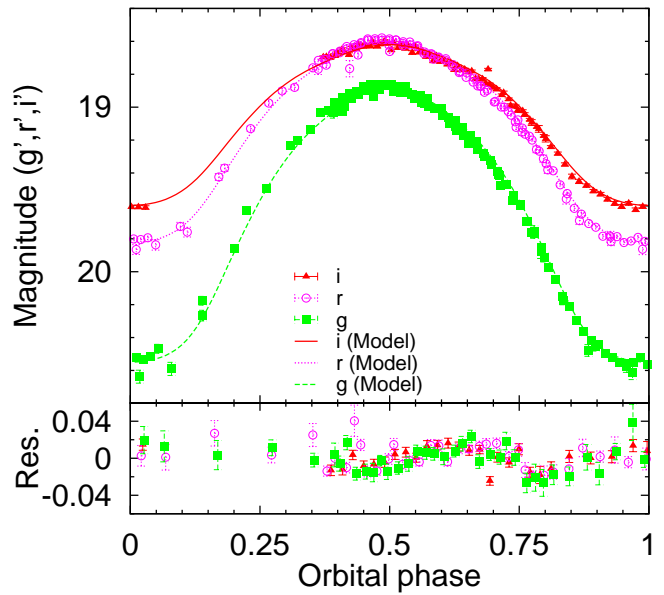}}}
  \caption{
Model fits to the orbital light curves in three bands ({\it top}) and
residuals ({\it bottom}).
} %
    \label{fig:LCmodel}
 \end{center}
\end{figure}

\begin{figure*}[ht!]
  \begin{center}
  \resizebox{2.0\columnwidth}{!}{\rotatebox{0}{\includegraphics[]{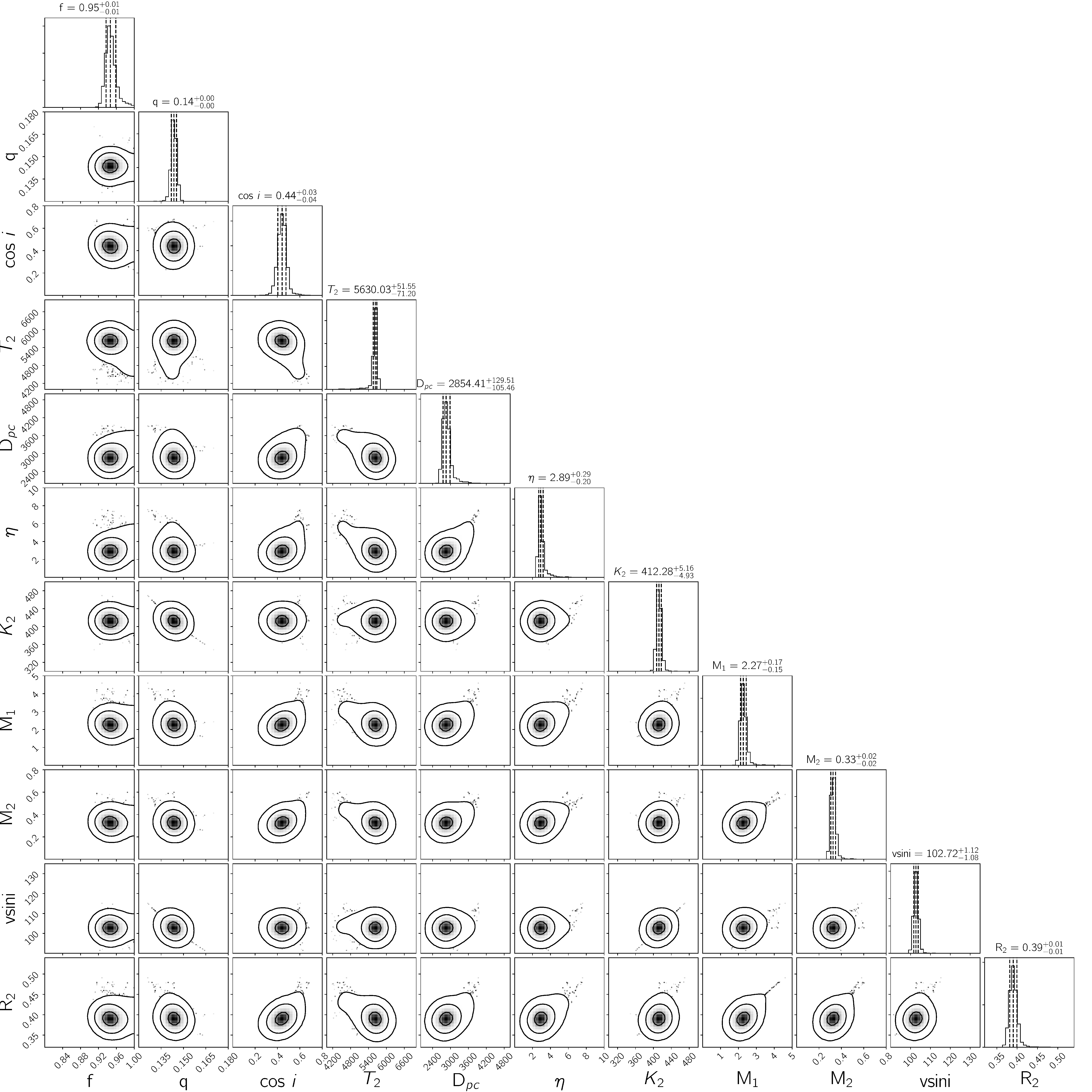}}}
  \caption{
Two-dimensional parameter distributions from our {\sc xrbcurve} fits
to the RVCs and LCs of J2215. Contours show the 1, 2, 3 and 4 sigma
confidence regions, and the right panels the projected one-dimensional
parameter distributions, together with the 1-sigma errors.
} %
    \label{fig:pspace}
 \end{center}
\end{figure*}

Using the full ensemble of lines, on the other hand, yields
intermediate values of the RVC semi-amplitude (green triangles in
Fig.~\ref{fig:k2}; K$_\mathrm{All}$=398.8$\pm$2.3~km~s$^{-1}$ for a G5
template).
We also find a clear systematic dependence with the template's {\it
  spectral type} when using the full spectral range: a monotonic
increase of K$_\mathrm{All}$ towards later (cooler) spectral types.
We conclude that, when measuring radial velocities in strongly
irradiated systems such as J2215, the spectral range and reference
spectra must be chosen carefully in order to measure the pulsar mass
accurately (see Section~\ref{sec:discussion} for further discussion).

The systemic velocities that we find are all in the 40--70~km~s$^{-1}$
range, showing only small changes with spectral range or template
spectral type.
Averaging the results for spectral types G0-K0 we find
G=49.0$\pm$2.5(stat)$\pm$8.0(syst)~km~s$^{-1}$ (where the statistical
and systematic errors correspond to the standard deviation of the full
range and of all three spectral ranges, respectively).
From the same sine fits we find an epoch of zero phase (companion at
inferior conjunction) of
T0=56976.9501$\pm$0.0003(stat)$\pm$0.0008(syst)~MJD (TDB),
which we use together with the radio pulsar timing P$_{\rm orb}$
(Sec.~\ref{sec:molly}) in order to compute orbital phases.

\newpage

\section{Modelling}
\label{sec:model}

In order to obtain the most reliable masses and orbital parameters, we
modelled simultaneously the photometric three-band LCs and
the Mg-triplet and Balmer-line RVCs, using the \textsc{xrbcurve}
model.
The model, described in the following, has been successfully applied
to LCs and RVCs of neutron star and black hole X-ray binaries
\citep[][for more details]{Shahbaz00,Shahbaz03,Shahbaz17}.

\subsection{Binary parameters and irradiation}
\label{sec:moddes}

\textsc{xrbcurve} includes a nearly Roche-lobe-filling secondary star
heated by high-energy photons from the compact object and an accretion
disk (not included in this case since no disk emission lines are detected).
The binary system's geometry is determined by the orbital inclination
$i$, the mass ratio $q=M_{\rm 2}/M_{\rm 1}$ (where $M_{\rm 1}$ and
$M_{\rm 2}$ are the masses of the neutron star and secondary star,
respectively), and the Roche lobe filling factor of the secondary
star, $f$.
The orbital period P$_\mathrm{orb}$, the radial velocity
semi-amplitude of the secondary star's barycenter K$_2$ and the
distance to the source in kpc (D) set the scale of the system.
The light arising from the secondary star depends on its mean,
unperturbed effective temperature (T$_{\rm 2}$) and the gravity
darkening exponent $\beta$.

The additional light due to irradiation is given by the irradiation
efficiency $\eta \equiv $L$_\mathrm{irr}/\dot{E}$, which we define in
accordance with previous work as the ratio of the heating luminosity
(L$_\mathrm{irr}$, assuming isotropic emission by an irradiating
point source at the compact object) to the spin-down luminosity of the
pulsar
\citep[$\dot{E}$=5.29$\times$10$^{34}$~erg~s$^{-1}$,][]{Breton13}.
We calculate the resulting increase in the local effective temperature
due to the irradiating external source assuming that all irradiating
flux is thermalized \citep{Shahbaz03}.
We use \textsc{nextgen} model-atmosphere fluxes \citep{Hauschildt99}
to determine the intensity distribution on the secondary star
and a quadratic limb-darkening law with coefficients taken from
\citet{Claret00}, to correct the intensity.
Based on the observed mid-G spectral type for the secondary star
(Sec.~\ref{sec:results}), we fix $\beta$ at 0.10 \citep{Lucy67}.

To model the RVCs, we assume that the whole secondary star contributes
to both Balmer and MgI-triplet lines, including the inner/irradiated
face.
This is based on the corresponding EWs, which follow those expected
from an A5--G5 star (Fig.~\ref{fig:ewsrc}).
We set the Balmer and MgI-triplet absorption line strengths according
to the effective temperature of each surface element of the star.
Using our template spectra, we determine the EW versus temperature
relationship for the same exact wavelength ranges used in measuring
the RVCs (Appendix~\ref{app:uvespop}; see Fig.~\ref{fig:ewtemp}).
The line strengths for each surface element on the star are then
calculated using its temperature and the EW-temperature relation.
Finally, radial velocities are calculated from the model line profile,
averaged among all surface elements visible at each orbital phase.

In determining the binary parameters, we use a Markov chain Monte
Carlo (MCMC) method convolved with a differential evolution fitting
algorithm \citep[DREAM; see][]{Shahbaz17,Vrugt16}. We use a Bayesian
framework to determine our binary model parameters
\citep{Gregory05}. We include the projected semi-major axis of the
pulsar orbit measured from radio timing observations (x$_1$ = a$_1$
sin\,i=0.46814[1] light-seconds; \citealt{Abdo13}) as an independent
constraint on q and K$_2$ (q=K$_2$ P$_\mathrm{orb}$ / 2$\pi$x$_1$c;
where c is the speed of light) using a Gaussian priori. Our MCMC
fitting makes use of flat prior probability distributions for the rest
of model parameters. We use 20 individual chains to explore the
parameter space and 40000 iterations per chain. We reject the first
500 iterations and only include every 10th point.

We use a reddening of E(B-V)=0.38~mag, which we calculate from the
X-ray absorbing column density towards J2215, N$_{\rm
  H}$=2.1$\times$10$^{21}$\,cm$^{-2}$~\citep{Gentile14} using the
conversion from \citep{Predehl95}.
Because N$_{\rm H}$ is in turn estimated from the measured pulsar
dispersion measure \citep{He13}, we verified the accuracy of the
reddening towards J2215 in different ways.
The total Galactic N$_{\rm H}$ in the direction of J2215 is 40\% higher
\citep{Kalberla05}, although this may include additional absorbing
material in the line of sight. Our value of E(B-V)=0.38 mag is
consistent with that measured from IR dust maps in the same direction
(0.35+/-0.02 mag, \citealt{Schlegel98}).
In order to quantify the possible impact on the measured pulsar mass,
we repeated the LC and RV fits with E(B-V) left as a free parameter,
with a flat priori. Reassuringly, this yields the same value
M$_1$$\simeq$2.3~M$_\odot$, and a value of E(B-V) fully consistent
with (less than one sigma from) the one we use.

We also compute the line-of-sight temperature at orbital phases 0.0
and 0.5 which represent the cold/night (T$_{\rm N}$) and hot/day
(T$_{\rm D}$) temperatures of the secondary star.
Using the spectral type measurements explained above, we impose
temperature constraints on T$_{\rm N}$ (5280--5920 K, corresponding to
a spectral type G5$\pm$5) and T$_{\rm D}$ (7800--8550 K, corresponding
to a spectral type A5$\pm$2).
Comparing with previous models of J2215 (Table~\ref{table:model}), we
conclude that our quantitative independent constraints on T$_{\rm N}$
and T$_{\rm D}$ are critical in order to find a good solution.
From our measured f$_{\rm veil}$$\simeq$0.8 (Sec.~\ref{sec:temp}), it
is clear that there is an extra light component that veils the
observed light from the secondary star.
To allow for this wavelength-dependent veiling we include an extra
flux component in the model LCs, $f_g$, $f_r$ and $f_i$ in the $g$,
$r$ and $i$ band, respectively.
We also allow for possible uncertainties in the absolute flux
calibration of the light curves, by including a wavelength-dependant
magnitude offset in the same bands.

The model parameters that determine the shape and amplitude of the
optical LCs, and RVCs are $i$, $T_{\rm 2}$, $\eta$, D, K$_2$ and the
extra flux components $f_g$, $f_r$ and $f_i$. There are also a number
of extra parameters: the phase shift for the LCs and RVCs (discussed below)
as well as the systemic velocities for the MgI-triplet and the
Balmer-line radial velocity curves (which we set to be the
same).
The $g',r',i'$-band LCs were phase folded
and averaged into 37, 37 and 28 orbital phase bins, respectively.
The MgI-triplet and Balmer-line RVCs contain 20 data points each
(Sec.~\ref{sec:molly} for details).
Given that there are five different data sets with different numbers
of data points, to optimize the fitting procedure we assigned relative
weights to them. After our initial search of the parameter space,
which resulted in a good solution, we scaled the uncertainties in each
data set (i.e. the LCs and the RVCs) so that the total reduced
$\chi^2$ of the fit was $\sim$1 for each data set separately. The MCMC
fitting procedure was repeated to produce the final set of parameters,
which were used to determine M$_1$ and M$_2$.

\subsection{Model results and comparison with previous work}
\label{sec:modres}

Our physical model fits simultaneously the Balmer and MgI RVCs of
J2215 (Figure~\ref{fig:rvcompare}, right) as well as the optical LCs
in three bands (Figure~\ref{fig:LCmodel}), with a global reduced
$\chi^2$ of 1.30 for 126 d.o.f.
Small residuals are apparent in the LC fits (especially near
$\Phi_{\rm orb}$=0.8) and in the RVC fits (at $\Phi_{\rm
  orb}$=0.5).
We show in Figure~\ref{fig:pspace} the parameter distributions from
our {\sc xrbcurve} fits to J2215.
All model parameters are well constrained, and the overall agreement
between the data and model is good.
Our best-fit D=2.9$\pm$0.1~kpc is fully consistent with the 3 kpc
value found independently from the MSP dispersion measure \citep{Hessels11}.

\begin{table}[ht]
  \scriptsize
\caption{Comparison of irradiated binary models for J2215. Ad-hoc
  constraints on the models are marked in italics.}
\centering
\begin{tabular}{l c c c c}
\hline\hline
Param. & This work\footnote{Our {\sc xrbcurve} fit to three-band photometric light curves (g, r and i) and two-band spectroscopic radial velocity curves (MgI and Balmer). See text for details.} & SH14\footnote{{\sc elc} fit to B, V and R lightcurves \citep[][and references therein]{Schroeder14}. Ranges from both their NextGen and PHOENIX models.} & R15\footnote{{\sc elc} fit to the same three-band BVR light curves and one radial velocity curve \citep{Romani15}. Their preferred 'HiT' model (no errors reported).} & R16\footnote{{\sc icarus} fit to the same BVR LCs and RVC, with irradiation from an extended intra-binary shock and temperature constraint, from \citet{Romani16}. Their preferred 'IBS-Td' model.} \\
\hline
$f$                 &  0.95$^{+0.01}_{-0.01}$  & 1.00$\pm$0.01 & $\simeq$1 &  0.905$\pm$0.004 \\
$q$          & 0.144$\pm$0.002 & 0.155--0.180 & 0.145 & ? \\
$i$($^\circ$)   &  63.9$^{+2.4}_{-2.7}$  & 49.5--54.3 & 88.8 & 83$\pm$6 \\
$D$(kpc) & 2.9$\pm$0.1 & n.a. & 3.9? & 3.0? \\
$K_{\rm 2}$(km/s) & 412.3$\pm$5.0 & 329--382\footnote{Predicted from LC model fit.} & 407.8\footnote{Derived from their best-fit q, i and $M_{\rm 1}$.} & ? \\
$\eta$       & 2.9$^{+0.3}_{-0.2}$  & 0.083$\pm$0.001 & 0.97 & 2.16$\pm$0.15 \\
$M_{\rm 1}$($M_{\odot}$) & 2.27$^{+0.17}_{-0.15}$ & 1.9--2.7 & 1.59 & ? \\
$M_{\rm 2}$($M_{\odot}$)  & 0.33$^{+0.02}_{-0.02}$ & 0.34--0.44 & 0.23 & ? \\
$T_{\rm 2}$ (K)  & 5630$^{+52}_{-71}$ & 3765--3945 & 6220 & ? \\
$T_{\rm N}$(K)  & {\it 5280--5920} & n.a. & {\it $>$6000} & 6416$\pm$58 \\
$T_{\rm D}$(K)  & {\it 7800--8550} & 4876--5090 & 9264\footnote{Derived from their best-fit M$_1$, M$_2$ and L$_{\rm irr}$.} & {\it 9000} \\
\hline
$\chi^2$/dof  &  164/126=1.30 & 1.5 & 4.4 & 1018/263=3.9 \\
\hline\hline
\end{tabular}
\label{table:model}
\end{table}

Table~\ref{table:model} presents our best-fit values and their 1-sigma
uncertainties, compared to previous studies of J2215.
First, as already pointed out by \citet{Romani15}, previous attempts
at determining orbital parameters exclusively based on photometric
measurements and modelling are unsuccessful (\citealt{Schroeder14};
see also \citealt{Breton13}).
This is clear, e.g., from the K$_2$ velocities predicted by those
photometric fit results, inconsistent with our measured values
(Table~\ref{table:model}).
Thus we conclude that, at least in the presence of strong irradiation,
dynamical information is required in order to find a reliable orbital
solution.

Second, the orbital inclination depends strongly on the temperatures
of both sides of the companion.
Indeed, a larger temperature contrast between both sides requires a
smaller inclination angle to produce the observed peak to peak
magnitude difference.
Our temperature constraints on the model are taken from quantitative
temperature measurements at $\Phi_{\rm orb}$=0.0 and 0.5
(Sec.~\ref{sec:temp}), and lead to a robustly determined
i=63.9$^\circ$$^{+2.4}_{-2.7}$.
Therefore, independent constraints on the temperature at different
orbital phases are also needed to find a robust solution.

Third, we have shown that a point-source irradiation binary model can
fit satisfactorily the J2215 data, in clear contrast with previous
results \citep{Romani15,Romani16}.
This discrepancy might be also due to the different temperature
constraints, but a more detailed comparison is warranted.
In any case, our results show that irradiation from an extended shock
is not required to explain the optical properties of J2215.

On the other hand, all four models do agree on the filling factor,
showing a nearly Roche-lobe filling companion in J2215.
We find an additional non-stellar flux contribution in the range
0.035--0.07~mJy in all three bands (g, r and i), between 2 and 10
times fainter than the companion star (at $\Phi_{\rm orb}$=0 and 0.5,
respectively).
This extra non-variable flux component, with a rather flat spectral
slope, might be due to synchrotron emission from an intrabinary shock,
but at present this interpretation remains tentative.
We can also rule out the presence of a quiescent disk \citep[as
  suggested by][]{Schroeder14}, based on the absence of broad hydrogen
and helium emission lines regularly associated with disks.

Our best-fit model predicts a projected rotational velocity for the
companion of Vsini=103$\pm$1~km~s$^{-1}$.
Applying an optimal subtraction (Sec.~\ref{sec:molly}) of the MgI
triplet line region around $\Phi_{\rm orb}$=0, with a G5 template
broadened in steps of 10~km~s$^{-1}$ up to 210~km~s$^{-1}$, we measure
Vsini=180$\pm$20~km~s$^{-1}$ (with a limb darkening coefficient u=0.8)
and Vsini=165$\pm$15~km~s$^{-1}$ (with no limb darkening, u=0).
Thus, taking into account the uncertainty on the amount of limb
darkening, our current observational constraints put Vsini in the
range 150--200~km~s$^{-1}$.
However, better spectral resolution spectra are required to measure
this accurately and compare it to our model prediction (our GTC data
have a resolution of 160~km~s$^{-1}$; Sec.~\ref{sec:spec}).
Finally, our best-fit value of $\eta$ implies a very high
L$_\mathrm{irr}$=[1.5$^{+0.3}_{-0.1}$]$\times$10$^{35}$~erg~s$^{-1}$.
An irradiating luminosity three times higher than the spin-down energy
budget might be explained by e.g. beaming/anisotropy of the pulsar
wind \citep[e.g.,][]{Philippov15}.

\begin{figure*}[]
  \begin{center}
  \resizebox{2.0\columnwidth}{!}{\rotatebox{0}{\includegraphics[]{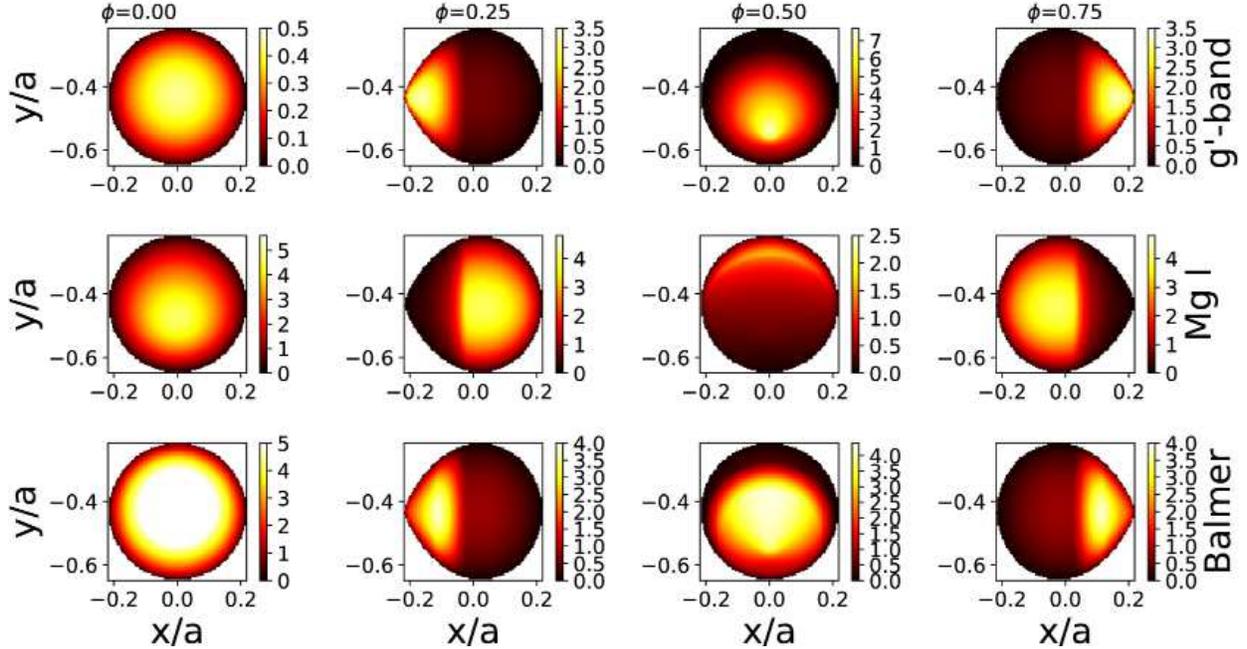}}}
  \caption{
Flux maps of the companion star in J2215 from our best-fit {\sc
  xrbcurve} model, seen at different orbital phases (as indicated
along the top axis). Flux units are arbitrary.
Three different sets of maps are presented, for: continuum g-band
emission ({\it top panels}), MgI triplet absorption lines ({\it
  middle}) and Balmer absorption lines ({\it bottom}).
The different distribution of lines throughout the surface is evident,
and due to the strong irradiation effects (see text for details).
The center of light depends strongly on the line or spectral
range choosen.
} %
    \label{fig:map}
 \end{center}
\end{figure*}

The flux distribution along the surface of the companion star is shown
in Figure~\ref{fig:map}, a by-product of our best-fit model of J2215.
These maps illustrate the drastic irradiation or 'heating' effects of
the pulsar wind and high-energy emission on the temperature
distribution of the secondary star.
Besides producing a strong temperature gradient between opposite sides
(which we measure in Sec.~\ref{sec:temp}), such strong heating shifts
the effective center of the secondary star (``center of light'')
away from its center of mass.
Because the strength of the lines varies throughout the surface of the
star, this shift is different for different absorption lines.
This in turn results in significant distortion of the integrated line
profiles as well as the corresponding RVCs
\citep[e.g.,][]{Phillips99,Shahbaz00}.
Thus our model provides a natural explanation for the systematic
difference in K velocities that we find and report in
Section~\ref{sec:rvcs}.

\subsection{On orbital phase shifts and systematics}
\label{sec:shift}

Our model allows for phase shifts in the LC and both RVCs, which may
arise physically from e.g. asymmetric heating of the companion star.
The best-fit values for the phase shifts in the Balmer RVC, the MgI
RVC and the LCs are, respectively: $\Delta\Phi_{\rm
  Balmer}$=0.018$\pm$0.003, $\Delta\Phi_{\rm MgI}$=0.0002$\pm$0.0039
(i.e., consistent with 0 within 1-sigma) and $\Delta\Phi_{\rm
  g}$=0.0082$\pm$0.0007 (where the quoted errors are again 1-sigma
statistical).
Similar LC phase shifts $\sim$0.01 have been reported for J2215
\citep{Schroeder14,Romani15,Romani16}, and interpreted as evidence
for asymmetric heating from an intra-binary shock.

However, since the model does not take into account errors on the
orbital phases, one must consider carefully the systematic uncertainty
on $\Phi_{\rm orb}$ before interpreting such phase shifts.
If we used the pulsar timing reference epoch to calculate $\Phi_{\rm
  orb}$ (which dates from 2009-12-21, from \citealt{Abdo13}), we
should include the orbital period derivative $\dot{P}_{\rm orb}$ (not
doing so would yield a propagated uncertainty to our 2014-08-02 epoch
of 0.02 orbital cycles).
But assuming a constant $\dot{P}_{\rm orb}$ over a 5-yr timespan is
problematic, since redbacks and black widows are known to show large,
erratic changes in $\dot{P}_{\rm orb}$ on timescales of years
\citep[e.g.,][]{Archibald13}.
Instead, we use our own spectroscopic reference time (T0, see
Section~\ref{sec:rvcs}) with a systematic uncertainty of 0.0008~d,
which corresponds to a systematic uncertainty on $\Phi_{\rm orb}$ of
$\delta\Phi_{\rm orb}$=0.0046 cycles.

In light of this, the only phase shift which we deem (marginally)
significant is $\Delta\Phi_{\rm Balmer}$, which is larger than 0 at
the 3.3-sigma confidence level (this shift can be seen in close
inspection of Figure~\ref{fig:rvcompare}).
Thus we conclude that, even if orbital phase shifts are necessary to
allow a good statistical fit to the data, their significance and
interpretation depend critically on the systematics of the orbital
ephemeris.
From our data and model of J2215, after quantifying the systematic
error on our orbital phases, we find one relatively small phase shift
in the Balmer-line RVC and thus only marginal evidence for heating
asymmetries.

\section{Discussion}
\label{sec:discussion}

\subsection{Measuring masses: An empirical K correction}
\label{sec:Kcorr}

In a careful study of the strongly irradiated companion star of
PSR~J2215+5135, we find that the radial velocities depend
systematically on the spectral features and spectral range used to
measure them (Sec.~\ref{sec:rvcs}).
Thanks to the high-quality GTC optical spectra, we show that magnesium
lines yield semi-amplitudes of the RVC that are
always higher than those inferred from hydrogen Balmer lines or from an
undetermined blend of lines.
We argue that this new systematic effect arises from the extreme
contrast between the the cold and heated sides of the companion star,
for which we measure temperatures of
T$_\mathrm{N}$=5660$^{+260}_{-380}$~K and
T$_\mathrm{D}$=8080$^{+470}_{-280}$~K, respectively
(Sec.~\ref{sec:temp}).
Under these circumstances, Balmer lines trace the hot face of the
companion star of J2215 in its orbit around the pulsar, while Mg-I
triplet lines trace its cold, unperturbed face.

We modelled both RVCs and the optical LCs in three bands, and found
the center of mass velocity to be K$_2$=412.3$\pm$5.0~km~s$^{-1}$
(Sec.~\ref{sec:model}).
Because the center of light of the Mg-I lines yields higher velocities
(K$_{Mg}$=K$_2$+$\Delta$K, $\Delta K > 0$), it is slightly shifted
{\it outwards} from the center of mass of the companion.
Using the simple expression for the center-of-light displacement
$\Delta$R/a=$\Delta$K/(K$_2$(1+q)) \citep{Wade88} and our best-fit
orbital parameters (Table~\ref{table:model}), we estimate $\Delta
R_{\rm MgI}/R_{\rm RL2} \simeq +0.11$ (i.e., about 11\% of the Roche
lobe radius of the companion).
Balmer lines, on the other hand, suffer from a stronger {\it inwards}
displacement of the center of light, which can yield to systematic
errors in dynamical mass measurements if not corrected.
We estimate this displacement as above using the measured
K$_{Balmer}$, and find $\Delta R_{\rm Balmer}/R_{\rm RL2} \simeq -
0.24$ (i.e., a 24\% shift relative to R$_{\rm RL2}$).

Given the large ($\simeq$10\%) differences in the measured K values
and the M$_1$$\propto$K$^3$ relation, this has important consequences
for the measured pulsar masses.
For instance, using the different measured K values and our best-fit
orbital solution
(Sec.~\ref{sec:model}) yields inconsistent values for the neutron star
mass: M$_1$=1.88$\pm$0.16~M$_\odot$ (from
K$_\mathrm{Balmer}$=382.8$\pm$4.7~km~s$^{-1}$) and
M$_1$=2.49$\pm$0.23~M$_\odot$ (from
K$_\mathrm{Mg}$=420.2$\pm$6.2~km~s$^{-1}$).
Thus we find that, in the presence of strong irradiation, the
systematic error on K may be equally or more important than the
uncertainty on the orbital inclination, i.


To circumvent this and reduce drastically the systematics on K
measurements, we have put forward a new method: we measure K
velocities using different sets of lines in order to ``bracket'' the
center-of-mass velocity of the companion star.
We have deemed as optimal the Balmer and MgI triplet lines
(Sec.~\ref{sec:molly}) which provide a lower and an upper limit on
K$_2$, respectively.
These could in principle be replaced by other sets of lines that trace
the movement of both the cold/dark and irradiated/bright sides of the
star.
While the traditional ``K correction'' relies on model assumptions or
on the transient nature of irradiation in some systems
(Sec.~\ref{sec:rvirr}), our method provides a direct way of
quantifying this correction from the same spectra of the irradiated
companion star.

To our knowledge, the only similar studies in the literature involve a
combination of emission and absorption lines in white dwarf binaries
\citep[e.g.,][]{Parsons10,Rodriguez-Gil15}.
Our method relies exclusively on absorption lines from the star's
atmosphere, thus eluding the uncertainty sometimes associated with the
exact site where Balmer/He/Bowen emission lines are formed.
This ``empirical K correction'' is of particular relevance for the
emerging population of MSPs in compact binaries, and it should also be
applicable in the broader context of semi-detached binaries with
strong irradiation/heating effects.

In summary, our findings show that metallic lines in general and MgI
lines in particular offer a much less distorted view of the center of
mass of the secondary star, opening a new way to measure masses in
strongly irradiated compact binary MSPs.
In the relevant temperature and spectral type range for J2215 and most
redback and blackwidow companions (spectral types A through M), Balmer
lines are more sensitive to temperature than Mg-I triplet lines (see
Appendix~\ref{app:uvespop}, Fig.~\ref{fig:ewtemp}).
Indeed, when going from a spectral type A5 to a G5, the EW of Balmer
lines decreases by a factor $\simeq$7 (48$\rightarrow$7~\AA), wheras
the EW of Mg-I triplet lines increases by a smaller factor $\simeq$3
(1.52$\rightarrow$4.75~\AA).
This may explain why the effects of irradiation are more drastic on
Balmer absorption lines than on metallic lines.

\subsection{A broader look at irradiation: K correction and deep heating}
\label{sec:rvirr}

Irradiation in compact binaries has been studied in the context of
black hole and neutron star low-mass X-ray binaries (LMXBs) as well as
white dwarf (WD) binaries (dwarf novae, DN, post-common envelope binaries,
PCEBs and asynchronous polars, AP).
In those systems, irradiation proceeds mostly through X-ray and UV
photons from a hot white dwarf or innermost accretion disk.
The ratio of the maximum irradiating flux at the companion's surface
(near the inner Lagrangian point) over the companion's intrinsic
unperturbed flux provides a good way to quantify the importance of
irradiation in close binaries: $f_{\rm irr}=\frac{L_{\rm
    irr}~R^2_2}{L_2~a^2}$.
The effects of irradiation on the measured K velocities, on the other
hand, are typically parameterized in terms of the so-called {\it K
  correction}, which we define as the ratio between observed and
center-of-mass K values: $f_{\rm K}=\frac{K_{\rm obs}}{K_2}$.
This correction is estimated in the literature in a number of
different ways, e.g., by comparing outburst and quiescence K values
\citep{Hessman84} or by simulating and fitting RVCs
with irradiation models \citep{Wade88,Phillips99,MunozDarias05}.

%
%
In some cases (compact LMXBs in outburst, very hot WDs),
chromospheric/fluorescent {\it emission} lines are formed on the {\it
  inner face} of the companion star, leading to a {\it lower limit} on
the center of mass velocity semi-amplitude K$_2$ (i.e., $f_{\rm K} <$1).
This is the case of Bowen fluorescence lines in LMXBs.
It is interesting to compare J2215 with the accreting millisecond
pulsars (AMPs) SAX~J1808.4--3658 and XTE~J1814--338, with Bowen-line
outburst K corrections of $f_{\rm K}$=0.90 and 0.81, respectively
\citep{Cornelisse09,Wang17}.
These are analogous to (and possibly evolutionary precursors of)
blackwidow and redback MSPs.
Assuming an irradiating luminosity in outburst L$_X \sim
10^{36}~erg~s^{-1}$, we estimate their maximum ratio of irradiating to
intrinsic flux and find extreme values, f$_{\rm
  irr}\sim$10$^3$--10$^4$.
For the persistent atoll sources 4U~1636--536 and 4U~1735--444, with
$f_{\rm K}\simeq$0.77 \citep{Casares06}, we estimate even higher
f$_{\rm irr}\sim$10$^5$.

Balmer/HeI emission lines in WD binaries are also associated with
strong, yet less extreme, irradiation. The AP V1500Cyg, for instance,
has an estimated $f_{\rm K}$=0.65 \citep[Balmer and HeII
  lines,][]{Horne89} and we find f$_{\rm irr}\sim$50 \citep[using the
  stellar parameters in][]{Schmidt95}.
Similarly, we find f$_{\rm irr}\sim$140 for the PCEB NNSer, with a
reported $f_{\rm K}$=0.88 \citep[Balmer lines,][]{Parsons10}.
This suggests that the irradiating flux must be at least ten times
higher than the intrinsic stellar flux in order to form emission lines
on the heated face.

In other cases (DNe and BHCs in outburst), {\it absorption} lines are
partially quenched on the inner face of the companion due to
irradiation, so that the effective center of light for these lines is
shifted towards the {\it outer face} and they provide an {\it upper
  limit} on K$_2$ (i.e., $f_{\rm K} >$1).
Even though stellar atmospheres with external heating are poorly
understood, the reduced absorption line strength is often attributed
to the reduced vertical temperature gradient in the presence of an
external UV/X-ray photon flux.
This is the case of the WD binaries ZCha, UGem and SSCyg, with $f_{\rm
  K}$ estimated at 1.03, 1.04 and 1.26, respectively
\citep{Wade88,Friend90,Hessman84}.
For these systems we find mild irradiation, with estimated $f_{\rm
  irr}$ values in the range 0.1--3.
The BHC GRO~J1655-40 on the other hand, with $f_{\rm K}\simeq$1.16,
has a relatively luminous F6IV companion star which also leads to a
mild f$_{\rm irr}\sim$7 \citep{Phillips99,Orosz97}.

In our case (J2215) and in compact binary MSPs in general, the
relativistic pulsar wind and gamma-ray emission are the dominant
sources of irradiation.
These feature typical spin-down luminosities
$\dot{E}$=10$^{34}$--10$^{35}$~erg~s$^{-1}$ and gamma-ray luminosities
(L$_\gamma$) about ten times lower.
Their X-ray luminosities are two to five orders of magnitude lower
than $\dot{E}$ and are thus less important in terms of irradiation
\citep[L$_\mathrm{X}$=10$^{30}$--10$^{32}$~erg~s$^{-1}$;][]{Linares14c,Gentile14}.
Indeed, J2215 has
L$_\mathrm{X}$=1.2$\times$10$^{32}$~erg~s$^{-1}$ (0.5--10~keV),
L$_\gamma$=1.4$\times$10$^{34}$~erg~s$^{-1}$ (0.1--100~GeV) and
$\dot{E}$=5.3$\times$10$^{34}$~erg~s$^{-1}$ \citep[][respectively; for
  a 2.9~kpc distance]{Linares14c,Acero15,Breton13}.
Our measured radial velocity amplitudes K$_{\rm Balmer}$ and K$_{\rm
  MgI}$, together with the center-of-mass velocity from our best-fit
model (K$_2$) imply K correction factors for J2215 of $f_{\rm
  K}$=0.928 and $f_{\rm K}$=1.019, respectively.
In other words, the K values measured using Balmer and MgI lines
  are 7.2\% lower and 1.9\% higher, respectively, than the true
  center-of-mass K$_2$.

For comparison, the K correction inferred by \citet{Kerkwijk11}
  for the black widow pulsar PSR B1957+20 was close to 8\%: $f_{\rm
    K}$=0.918 (where their $K_{\rm obs}$ was measured using a slightly
  wider spectral range and a G2 template). Thus, K corrections may be
  similarly important in black widow and redback binaries. This can be
  understood qualitatively from the two terms entering $f_{\rm irr}$,
  which have opposite trends: $R^2_2/a^2$ is lower in black widows
  (lower irradiating flux due to smaller solid angle), but $L_{\rm
    irr}/L_2$ is higher compared to redbacks (as the black widow
  companions are less luminous).

Using our orbital solution (which gives R$_{\rm RL2}$/a=0.23),
our best-fit value of L$_{\rm irr} \simeq$3$\times$$\dot{E}$,
and a companion luminosity L$_2 \simeq
$5.3$\times$10$^{32}$~erg~s$^{-1}$ (from our best-fit T$_2$=5630~K and
R$_2$=0.39~R$_\odot$),
we estimate f$_{\rm irr}\sim$15.
In other words, the irradiating flux at the companion's heated face is
up to 15 times higher than the intrinsic stellar flux.
Compared with UV and X-ray photons, Gamma-ray photons and relativistic
particles from the pulsar wind are expected to penetrate deeper into
the companion atmosphere.
This leads to deep/internal heating of the companion's inner face, so
it is not surprising to find no emission lines in J2215, and no
quenching of absorption lines either (Fig.~\ref{fig:ewsrc}).

\begin{figure}[h]
  \begin{center}
  \resizebox{1.0\columnwidth}{!}{\rotatebox{-90}{\includegraphics[]{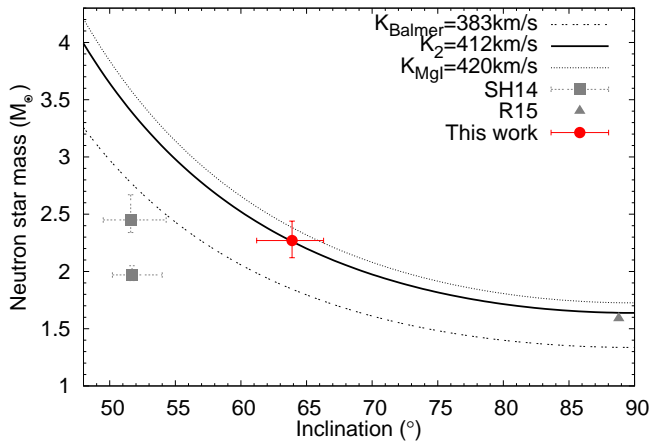}}}
  \caption{
    Neutron star mass measurements in J2215, shown vs. orbital
    inclination. Dotted and dashed lines show the effects of using
    different sets of absorption lines to measure velocities
    (Secs.~\ref{sec:rvcs} and \ref{sec:Kcorr}). The thick solid line
    shows our model K$_2$ (Sec.\ref{sec:model}). Our best-fit orbital
    solution (filled red circle) is compared with previous work
    (\citealt{Schroeder14,Romani15}; filled squares and triangles,
    respectively).
} %
    \label{fig:m1}
 \end{center}
\end{figure}

\subsection{The mass of PSR~J2215+5135}
\label{sec:mass}

We have argued that our ``empirical K correction'' removes a
critical systematic uncertainty in irradiated systems: the difference
between center of light and center of mass of the companion.
In J2215, our K$_2$=412.3$\pm$5.0~km~s$^{-1}$ yields a mass function
of 1.2~M$_\odot$, an absolute lower limit on the pulsar mass.
Combined with the tight constraints on q (Sec.~\ref{sec:model};
\citealt{Abdo13}), this implies a minimum neutron star mass of
1.6~M$_\odot$.
Thus we find that J2215 contains a neutron star more massive than the
``canonical'' 1.4~M$_\odot$ double neutron stars, adding to the
growing number of such systems
\citep[][]{Demorest10,Antoniadis13,Ozel16}.

Furthermore, we have shown that J2215 could harbor the most massive
pulsar known to date.
Our best-fit model yields a very massive neutron star with $M_{\rm
  1}$=2.27$^{+0.17}_{-0.15}$~M$_\odot$, and a well-constrained
inclination of i=63.9$^{+2.4}_{-2.7}$ (see Figure~\ref{fig:m1}).
This is more massive than the previous well-established record holder
\citep[2.01~M$_\odot$,][]{Antoniadis13}, at the 97\% confidence level.
The results of \citet{Romani15,Romani16} are clearly at odds, although
their nearly edge-on model (with a neutron star mass of 1.6~M$_\odot$)
has no uncertainties reported and admittedly fails to describe the
data.
  In a previous study of the original black widow pulsar,
\citet{Kerkwijk11} found a pulsar mass similarly high
(2.40$\pm$0.12~M$_\odot$).
The main advances of the method we have presented here are: i) an {\it
  empirical} K correction, based on radial velocities measured with
two different sets of lines (Sec.~\ref{sec:rvcs}); and ii) {\it
  independent} constraints on the temperature imposed on the model,
based on absorption line strengths (Sec.~\ref{sec:temp}).
We argue that these two new advances make our results more robust
compared to previous work \citep{Kerkwijk11,Romani15,Romani16}.
There may still be unknown or highly uncertain systematic effects,
however, biasing the best-fit model inclination and thus the dynamical
mass measurements \citep[see, e.g., the discussion in][their Section
  4]{Kerkwijk11}.

  We therefore conclude that, if confirmed with an independent
measurement of the orbital inclination, the massive neutron star in
J2215 may place new constraints on the equation of state at
supra-nuclear densities.
This would push the limits of the most massive neutron stars in our
Galaxy, setting a lower limit of 2.3~M$_\odot$ to their maximum
mass. Since particle interactions in the core provide the pressure
necessary to halt its collapse, the maximum mass of a neutron star
places independent constraints on how these particles interact
\citep{Lattimer07}. For instance, exotic forms of matter such as
hyperons or deconfined quarks have been proposed to exist in the
central parts of a neutron star, yet they can hardly account for a
neutron star as massive as the one we find in J2215 \citep[see
  also][]{Ozel16}.

With new Galactic MSPs being currently discovered at a rate of 10-30
per year \citep{LorimerCat}, the neutron star mass range will be
explored further in the next decade, and is likely to continue
widening. We have shown here that in the study of strongly irradiated
pulsar companions, a controlled measurement of temperatures and
velocities throughout the orbit is possible with current instruments,
and key to finding a robust dynamical solution. Our novel technique,
which combines velocity measurements with different absorption lines,
temperature measurements and physical modeling of the binary, should
provide a path forward for dynamical mass measurements in this growing
population.

\section{Summary and Conclusions}
\label{sec:conclusions}

\begin{itemize}[leftmargin=0.2cm]

  \item We have identified, for the first time and thanks to GTC's
    large collecting area, absorption lines from both sides of the
    irradiated companion star to PSR~J2215+5135.
We show that Mg-I triplet lines effectively trace the unheated ``dark
side'' of the companion, while hydrogen Balmer lines trace its
irradiated side.
We are therefore able to bracket the center of mass velocity, placing
both an upper and a lower limit on K$_2$.
This removes the systematic uncertainty on K$_2$ in strongly
irradiated systems due to the displacement of the center of light,
traditionally incorporated in the so-called ``K correction''.

  \item We argue that, beyond light curve modelling, accurate mass
    measurements in strongly irradiated binary systems require i)
    radial velocities, preferrably measured using metallic lines; and
    ii) robust constraints on the temperatures of both sides of the
    companion.

  \item In particular, we find that the semi-amplitude of the radial
    velocity curve of J2215 measured with MgI lines is systematically
    higher than that measured with Balmer lines, by 10\%.

  \item We measure temperatures for the cold and hot sides of
    T$_\mathrm{N}$=5660$^{+260}_{-380}$~K and
    T$_\mathrm{D}$=8080$^{+470}_{-280}$~K, respectively.
    
  \item By modeling jointly both radial velocity curves and the light
    curves in three bands, while imposing the temperature constraints
    above, we find that J2215 has: i) a center-of-mass K velocity of
    K$_2$=412.3$\pm$5.0~km~s$^{-1}$; ii) an inclination
    i=63.9$^\circ$$^{+2.4}_{-2.7}$; iii) an apparent irradiating
    luminosity three times higher than its spin-down luminosity and
    iv) a companion close to filling its Roche lobe (filling factor
    0.95$^{+0.01}_{-0.01}$).

  \item Our physical modeling can reproduce the measured fluxes and
      velocities without invoking extended irradiation, and yields
      only marginal evidence for asymmetric heating (in the form of
      orbital phase shifts).

  \item We thereby find that J2215 hosts a main-sequence G5 companion
    with M$_{\rm 2}$=0.33$^{+0.03}_{-0.02}$~M$_\odot$ and a very
    massive neutron star with $M_{\rm
      1}$=2.27$^{+0.17}_{-0.15}$~M$_\odot$.

    \item Pending independent confirmation of the orbital inclination,
      our results strongly suggest that the maximum neutron star mass
      is at least $\sim$2.3~M$_\odot$.
    
\end{itemize}

\textbf{Acknowledgments:}

Based on observations made with the GTC, WHT and IAC-80 telescopes
operated by the IAC and ING in the Spanish Observatories of el Roque
de los Muchachos (La Palma) and el Teide (Tenerife) under regular,
service and DDT programs.
We thank C. Fari{\~n}a, P. Chinchilla, R. Karjalainen,
A. Cabrera-Lavers, A. Oscod and R. Corradi for support during these
observations, and the IAC director R. Rebolo for granting the DDT.
This work was supported by the Spanish Ministry of Economy and
Competitiveness (MINECO) under the grant AYA2013-42627.
{\sc iraf} is distributed by the NOAO, operated by AURA under cooperative
agreement with NSF.
We thank T. Marsh for the use of {\sc molly} and {\sc ultracam}, and
acknowledge the use of data from the UVES Paranal Observatory Project
(ESO DDT Program ID 266.D-5655).
We appreciate discussions with R. Alonso, R. Breton, P. Charles,
J. Jos{\'e}, T. Mu{\~n}oz-Darias and P. Rodr{\'i}guez-Gil during
different stages of this work.
M.L. is supported by EU's Horizon 2020 programme through a Marie
Sklodowska-Curie Fellowship (grant nr. 702638).

\appendix

\section{A stellar spectral library for temperature and
  radial velocity measurements}
\label{app:uvespop}

\begin{figure}[h]
  \begin{center}
  \resizebox{\columnwidth}{!}{\rotatebox{0}{\includegraphics[]{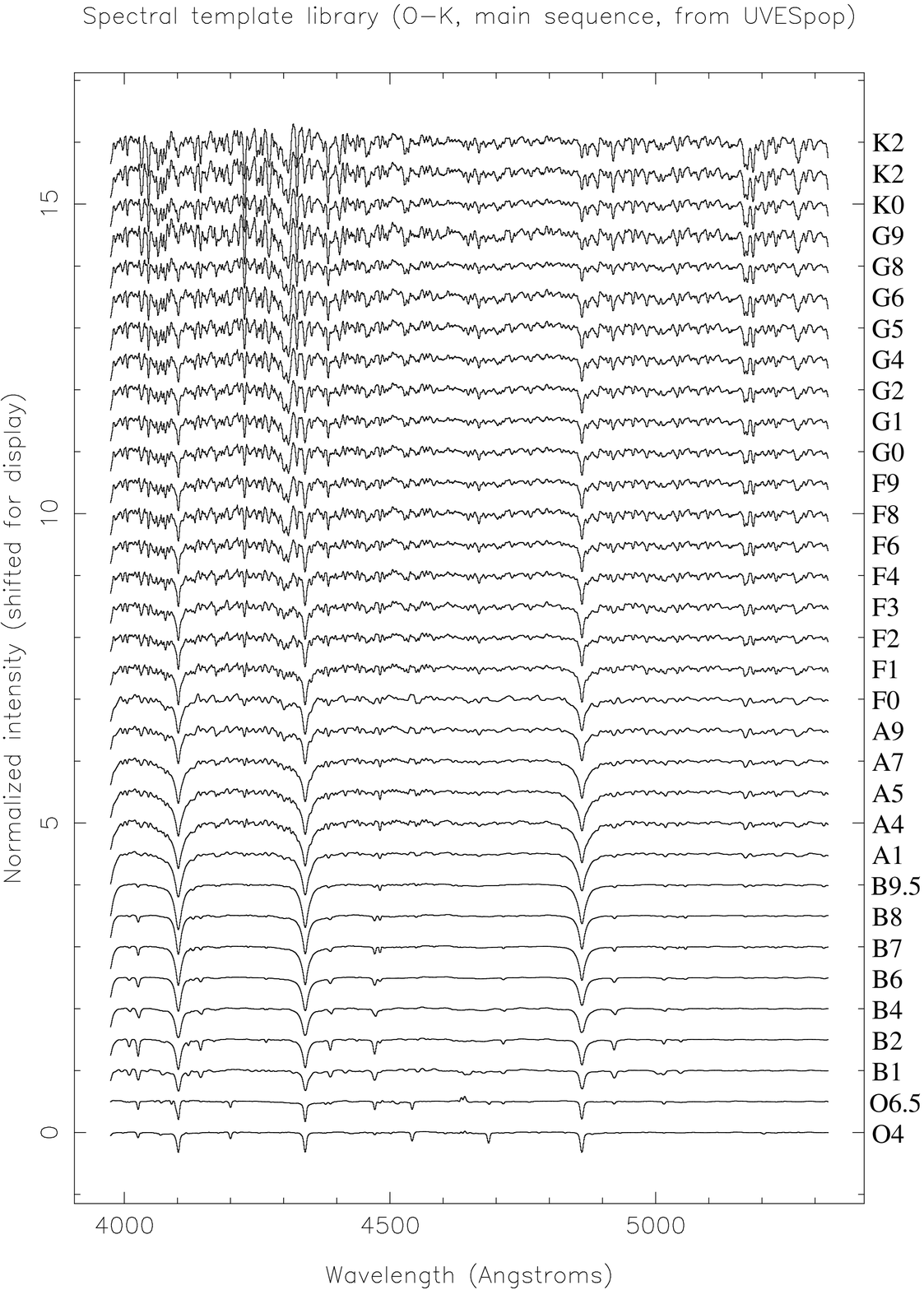}}}
  \caption{
Template library adapted from UVES-POP \citep{Bagnulo03} for
temperature and radial velocity measurements.
Normalized intensities are shifted for display, and the spectral type
is noted along the right axis.
\vspace{0.3cm}
} %
    \label{fig:uvespop}
%
  \resizebox{\columnwidth}{!}{\rotatebox{-90}{\includegraphics[]{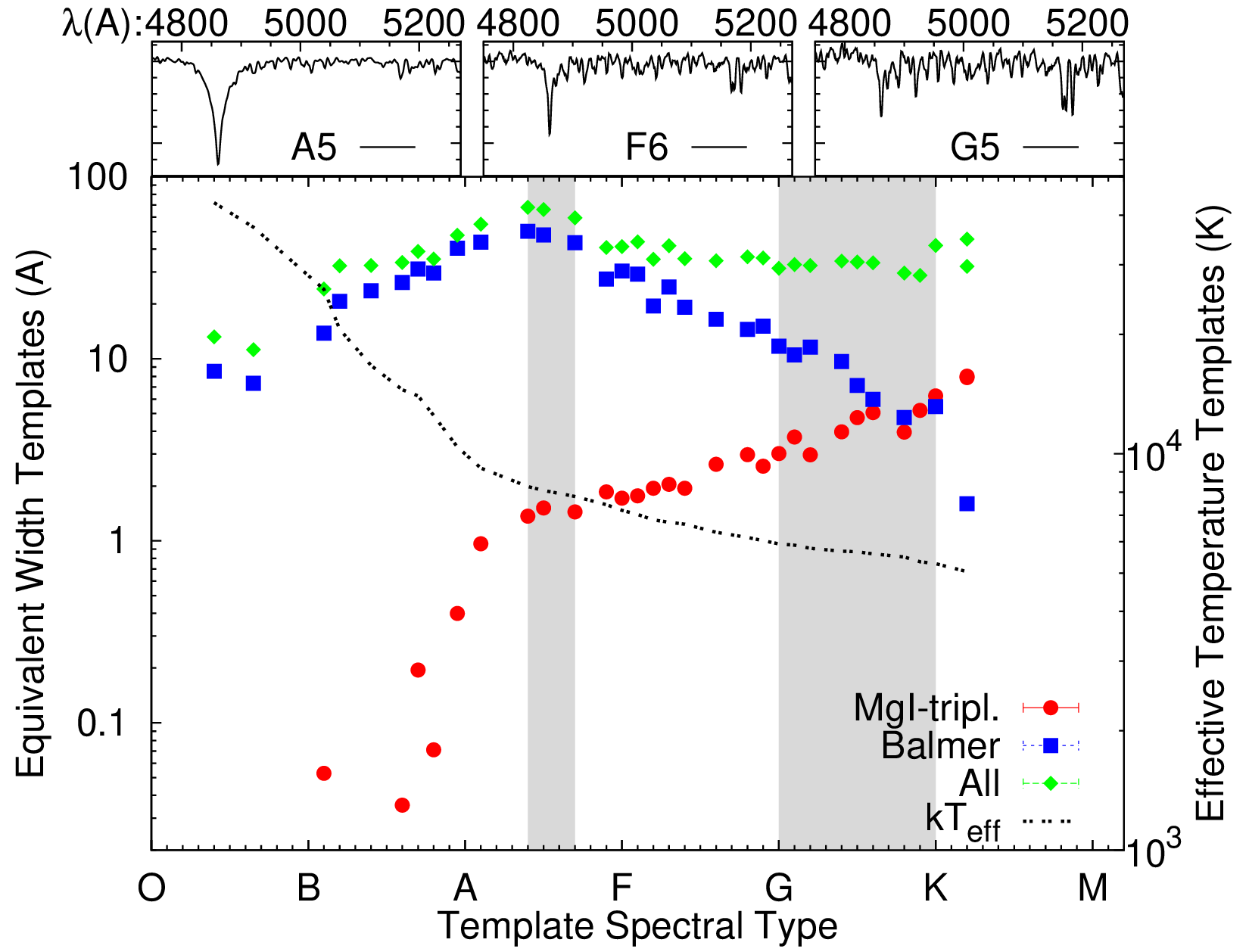}}}
  \caption{
EW of absorption lines from our template spectra, in the three ranges
used for radial velocity measurements: i) Balmer lines (blue squares);
ii) MgI triplet (red circles) and the full GTC-OSIRIS range (green
triangles).
The dotted line shows the template temperature (right vertical axis;
from \citealt{Pecaut13}).
Three representative spectra are shown in the top panels, within the
H$\beta$ and MgI-triplet line region.
} %
    \label{fig:ewtemp}
 \end{center}
\end{figure}

We built library of main sequence stellar spectra (or ``templates'')
in order to measure spectral types (temperatures) and radial
velocities (Section~\ref{sec:molly}).
We used 33 VLT-UVES spectra from the Paranal Observatory Project
\citep[UVES-POP]{Bagnulo03}, initially rebinned to a
0.2~\AA\ resolution, which cover the $\sim$3000--10000~\AA\ range.
We normalized them to their continuum level by fitting a spline
function, and excluded the gaps between echelle orders.

We then subtracted each template's radial velocity, measured by
cross-correlating the H$\alpha$ and H$\beta$ line profiles with a
Gaussian with FWHM=100~km~s$^{-1}$, or two such Gaussians separated by
200~km~s$^{-1}$ in the broad line cases.
These radial velocities were in good agreement with the values listed
in the SIMBAD database.
The resulting spectra are shown in Figure~\ref{fig:uvespop}, broadened
to match the spectral resolution of the GTC-OSIRIS spectra presented
herein (160~km~s$^{-1}$; Sec.~\ref{sec:spec}).
We also calculated the EW of absorption lines in the templates, shown
in Figure~\ref{fig:ewtemp}, which we use in our modelling of the
Balmer line and MgI-triplet radial velocities
(Section~\ref{sec:model}).

\bibliographystyle{apj}
\bibliography{../biblio.bib}

\end{document}